\documentclass[preprint2]{aastex}
\usepackage{rotating}

\newcommand\calh{{\cal H}}

\newcommand\Eg{E_{\rm g}}
\newcommand\Ek{E_{\rm k}}

\newcommand\kms{\mbox{\,km s}^{-1}}

\newcommand\Ncl{N_{{\rm cl}}}
\newcommand\nth{n_{{\rm th}}}

\newcommand\Msun{M_{\odot}}

\newcommand\pcc{\mbox{\,cm}^{-3}}
\newcommand\rt{RUN03}
\newcommand\rv{RUN20}
\newcommand\sigmav{\sigma_{\rm v}}
\newcommand\tff{\tau_{\rm ff}}

\newcommand\nav{\langle n \rangle}

\slugcomment{Submitted to ApJ}
\shorttitle{Clumps in simulations of collapsing molecular clouds}
\shortauthors{Camacho et al.}

\begin{document}

\title{Energy budget of forming clumps in numerical simulations of collapsing clouds} 
  
\author{Vianey Camacho$^1$, Enrique V\'azquez-Semadeni$^1$, Javier
Ballesteros-Paredes$^1$, Gilberto C. G\'omez$^1$, S. Michael Fall$^2$,
M. Dolores Mata-Ch\'avez$^1$}

\affil{1. Instituto de Radioastronom\'ia y Astrof\'isica, \\
      Universidad Nacional Aut\'onoma de M\'exico, Campus Morelia \\
      Apartado Postal 3-72, 58090, Morelia, Michoac\'an, M\'exico \\
      2. Space Telescope Science Institute \\
      3700 San Martin Drive \\
      Baltimore, MD 21218 US}


\begin{abstract}

We analyze the physical properties and energy balance of density
enhancements in two SPH simulations of the formation, evolution, and
collapse of giant molecular clouds. In the simulations, no feedback is
included, so all motions are due either to the initial, decaying
turbulence, or to gravitational contraction. We define clumps as
connected regions above a series of density thresholds. The resulting
full set of clumps follows the generalized energy-equipartition
relation $\sigma_{v}/R^{1/2} \propto \Sigma^{1/2}$, where $\sigma_{v}$
is the velocity dispersion, $R$ is the `radius", and $\Sigma$ is the
column density. We interpret this as a natural consequence of
gravitational contraction at all scales, rather than virial
equilibrium. Nevertheless, clumps with low $\Sigma$ tend to show a
large scatter around equipartition. In more than half of
the cases, this scatter is dominated by external turbulent
compressions that {\it assemble} the clumps, rather than by
small-scale random motions that would disperse them. The other half
does actually disperse. Moreover, clump sub-samples selected by means
of different criteria exhibit different scalings. Sub-samples with
narrow $\Sigma$ ranges follow Larson-like relations, although
characterized by their respective value of $\Sigma$.  Finally, we find
that: i) clumps lying in filaments tend to appear sub-virial; ii)
high-density cores ($n \ge 10^5$ cm$^3$) that exhibit moderate kinetic
energy excesses often contain sink (``stellar'') particles, and the
excess disappears when the stellar mass is taken into account in the
energy balance; iii) cores with kinetic energy excess but no stellar
particles are truly in a state of dispersal.

\end{abstract}



\section{Introduction}
\label{sec:intro}

Ever since the pioneering work of \citet{Larson1981} it has been 
recognized that molecular clouds (MCs) obey scaling relationships, 
that have been interpreted as representative of approximate virial 
equilibrium in the clouds between their internal ``turbulent'' motions
and their self-gravity. Subsequently, however, there have been 
suggestions that these relations may actually be the result of 
observational selection effects \citep[e.g.,] [] {Kegel1989, Scalo90}. 
In addition, there have been attempts at generalization of these 
relations \citep[e.g.,] [] {KM86, Heyer2009} and reinterpretations in 
terms of global cloud collapse rather than virialization 
\citep[][hereafter B11] {Ballesteros2011}. Moreover, there are 
structures that are observed to possess kinetic energies in excess of 
those that would be consistent with equilibrium (or more generally, 
energy equipartition).

In this paper we aim to investigate whether clumps forming in numerical
simulations of clouds undergoing global gravitational contraction exhibit 
similar properties as those in observational surveys such as Larson 
scaling relations, and search for a cause of the apparent kinetic energy
excesses seen in some subsets of clumps in observational samples.

\subsection{Larson's relations and their generalization} \label{sec:Larson_gen}

For over three decades, it has been accepted that molecular clouds (MCs)
satisfy the so-called \citet{Larson1981} scaling relations between
velocity dispersion ($\sigmav$), mean number density ($\nav$) and 
size ($L$). In their presently accepted form, these relations are 
\citep[e.g.,][] {Solomon+87, HB04}
\begin{equation}
	\nav \approx 3400 \left(\frac{L}{1 {\rm \, pc}}\right)^{-1} \pcc,
\label{eq:larson_dens}
\end{equation}
and
\begin{equation}
 	\sigmav \approx 1 \left(\frac{L}{1 {\rm \, pc}}\right)^{1/2} \kms.
\label{eq:larson_vel}
\end{equation}
\citet{Larson1981} additionally showed that these relations implied 
that the velocity dispersion is close to the value corresponding to virial
equilibrium. In what follows, we will more generally refer to this as
``near equipartition'' between the nonthermal kinetic and the 
gravitational energies. Also, it should be remarked that eq.\
(\ref{eq:larson_dens}) implies that the column density of the clouds,
$\Sigma = \int_{\rm LOS} \rho d\ell$ is approximately the same for MCs
of all sizes. In this expression, $\ell$ is the length element, and the
integration is performed along the line-of-sight (LOS) through the cloud.
  
However, the validity of Larson's relations has been questioned by 
various authors. \citet{Kegel1989} and \citet{Scalo90} argued that the 
apparent constancy of the column density may arise from selection 
effects caused by the need to exceed a certain minimum column density 
in order to detect the clouds, and by a maximum apparent column 
density caused by line saturation (optical thickening).
This possibility was in fact recognized by \citet{Larson1981} himself. 
Some time later, \citet{Ballesteros2002} showed that, in numerical 
simulations of turbulent clouds, clumps defined by means of a column 
density threshold exhibited a Larson-like density-size relation, but 
clumps defined by means of a {\it volume} density threshold did not.

Several years later, using the Boston University-FCRAO Galactic 
Ring Survey \citep[]{Jackson2006}, \citet[][hereafter H09]{Heyer2009}
re-analyzed the giant molecular cloud (GMC) sample of \citet{Solomon+87}. 
The higher angular sampling rate and resolution available to H09, as well 
as the use of the $^{13}$CO $J=1$--0 line allowed them to obtain a much 
larger dynamic range in column density than that available to 
\citet{Solomon+87}. Moreover, H09 considered two different definitions 
for the cloud boundaries, thus effectively obtaining two different 
MC samples.\footnote{Contrary to some claims in the literature, these 
two definitions of the cloud boundaries do amount to {\it two} different 
MC samples, as the masses and velocity dispersions were measured for 
each cloud within each of the two boundaries. Thus, the `A2' clouds in 
H09 constitute a sample of smaller, denser objects within the `A1' 
sample, just like dense clumps and cores are substructures of  their 
parent MCs, with independent dynamical indicators.} With this 
procedure, the GMC sample of H09 spanned over two orders of 
magnitude in column density, making it clear that column density is {\it 
not} constant for GMCs \citep[see also][]{Heyer+01}. Nevertheless, 
H09 noted that, in spite of the non-constancy of the column density, the 
GMCs are still consistent with virial equilibrium. They showed this by 
noting that their GMC sample satisfied
\begin{equation}
   \frac{\sigmav} {R^{1/2}} \approx \left(\frac{\pi G \Sigma} {5}\right)^{1/2}. 
\label{eq:gener_Larson}
\end{equation}
When the column density is not constant, this relationship corresponds
to virial equilibrium; i.e., to $|\Eg| = 2 \Ek$, with $\Ek$ being the
nonthermal kinetic energy and $\Eg$ being the gravitational energy 
for a spherical cloud of uniform density and radius $R$. Thus, eq.\
(\ref{eq:gener_Larson}) can be considered as the generalization of
Larson's relations when $\Sigma$ is not constant.

Shortly thereafter, \citet{Lombardi+10} claim\-ed that the column density
of GMCs is constant after all. Using near-infrared excess techniques, 
these authors argued that the mean GMC column density in their sample
remained constant in spite of being sensitive to very low extinctions, 
thus suggesting that the minimum-column density imposed by a 
sensitivity threshold was not an issue. 
However, it has subsequently been recognized that this effect is 
natural for clouds with a $\Sigma$ probability density function (PDF) 
that peaks at some value and drops fast enough at lower column 
densities \citep{Beaumont+12, Ballesteros2012}. 
The lack of pixels at low $\Sigma$ implies that the dominant apparent 
column density will be that of the peak, and it is now recognized that the 
presence of a peak may be an artifact of incomplete sampling at low 
column densities 
\citep{Lombardi+15}. Therefore, at present there is no compelling 
evidence for the validity of the density-size relation 
(eq.\ \ref{eq:larson_dens}) for GMCs nor their substructures in general.

On the other hand, the velocity dispersion-size relation,\footnote{In 
what follows, we will refer to this relation as the linewidth-size relation 
as well.} expression (\ref{eq:larson_vel}) above, has often been 
interpreted as the signature of supersonic turbulence, with an energy 
spectrum $E(k) \propto k^{-2}$, where $k$ is the wavenumber.
Indeed, the velocity variance, interpreted as the average turbulent 
kinetic energy per unit mass in scales of size $\ell \le 2\pi/k$, given by 
$\sigmav^2(\ell) = \int_{k>2 \pi/\ell} E(k) dk$, scales as $\ell^{1/2}$ 
\citep[e.g.] [] {VS+00a, ES04, MO07}. In this case, the velocity dispersion-
size relation would have a completely independent origin from that of 
the density-size relation, and the reason for the observed near-
equipartition between the gravitational and turbulent kinetic energies 
would require a separate explanation.  However, massive star-forming 
clumps notoriously do not conform to the $\sigmav$-$L$ relation 
\citep[e.g.,] [] {CM95, Plume+97, Shirley+03, Gibson+09, Wu+10}, 
a situation that appears inconsistent with a universal turbulent energy 
cascade spanning the whole range from the scale of GMCs down to the 
scale of massive clumps.

An alternative interpretation was suggested by \citet[] [hereafter B11]
{Ballesteros2011}, who proposed that the origin of the $\sigmav$-$L$
relation was not turbulence, but rather gravitational contraction of the
clouds, combined with the observational selection effect of a restricted
column density range.
This possibility was actually suggested over four decades ago by 
\citet{GK74}. Similarly, \citet{Liszt+74} suggested that their line profiles 
and LOS-velocity maps of the Orion MC were consistent with extended 
radial motions, although they could not discriminate between expansion 
and collapse. However, the extended-motion scenario was soon 
dismissed by \citet{ZP74}, who argued that, if that were the case, then 
the star formation rate in MCs should be much larger than observed, and 
that systematic shifts between emission lines produced by HII regions at 
the centers of the clouds and absorption lines produced in the radially-
moving cloud envelopes should be observed, but they are not. 
\citet{ZE74} then proposed that the observed linewidths corresponded 
to supersonic, small-scale turbulence.

The small-scale turbulence scenario, however, suffers from a number of
problems \citep[see] [for a detailed discussion] {VS15}.  Instead, B11 
have suggested a return to the scenario of gravitational collapse at the 
scale of the whole GMCs, with the problem of an excessive SFR being 
solved by early destruction of the clouds by stellar feedback \citep[][]{VS+10, ZA+12, Dale+12, Colin+13, ZV14}.  B11 noted that the generalized Larson relation, eq.\ (\ref{eq:gener_Larson}) 
is not only satisfied by GMCs, but also by massive clumps that do not 
satisfy Larson's velocity dispersion-size relation, eq.\ (\ref{eq:larson_vel}).
 
Thus, B11 interpreted the near-equipartition as evidence for free-fall in 
the clouds \citep[see also][]{Traficante+15} rather than near-virial 
equilibrium, noting that the virial and free-fall velocities differ only by a 
factor of $\sqrt{2}$.
Indeed, for a freely collapsing cloud, defining the total energy as zero, 
the nonthermal kinetic energy and the gravitational energy satisfy $\Ek 
=|\Eg|$,  so that, instead of eq.\ (\ref{eq:gener_Larson}), we have
\begin{equation}
        \frac{\sigmav} {R^{1/2}} \approx \left(\frac{2 \pi G \Sigma}
        {5}\right)^{1/2}. 
\label{eq:free-fall}
\end{equation}
Generally, the observational errors and uncertainties in cloud and clump
surveys are larger than this slight $\sqrt{2}$ factor, so that, for all 
practical purposes, {\it any evidence in favor of virial equilibrium based 
on energetics of the clouds can just as well be interpreted as evidence in 
favor of free collapse.}
Recent observational studies have shown signatures of infall 
motions in line profiles along filaments and massive clumps in the 
Cygnus X region \citep[][]{Schneider+10}, in massive star-forming cores 
of the infrared dark cloud SDC335.579-0.272 \citep[][]{Peretto+13}, and 
in massive starless cores \citep[][]{Traficante+15}, supporting the 
notion that this systems are consistent with a global gravitational 
collapse. Moreover, \citet{Traficante+15} introduce an equivalent 
analysis as in H09 to demonstrate that most of the non-thermal motions 
in their sample originate from self-gravity.

\subsection{Deviations from energy equipartition}
\label{sec:attempts}

An additional important feature in the GMC sample studied by H09 is that
the clouds tended to lie systematically {\it above} the virial-equilibrium
line in a plot of $\sigmav/R^{1/2}$ {\it vs.} $\Sigma$, a plot that we 
will refer to as the Keto-Heyer or KH diagram \citep{KM86, Heyer2009}.  
This feature has received different interpretations by  
different authors. H09 themselves interpreted it simply as a systematic 
underestimation of the cloud masses, due to the various assumptions they 
used in determining the masses from $^{13}$CO emission. On the other 
hand,  \citet{Dobbs2011} have interpreted this feature as evidence that 
most of the GMCs are gravitationally unbound, probably because they form 
by cloud-cloud collisions, which feed a large velocity dispersion that 
unbinds the GMCs. 
\citet{KM86} and \citet{Field2011}, instead, have assumed that the clouds 
are gravitationally unbound, but confined by an external pressure, while 
B11 suggested that the GMCs are actually collapsing, and that, at face value, 
the H09 data are slightly more consistent with free-fall than with virial 
equilibrium, since the free-fall velocity is slightly larger than the virial one. 

In addition to the slight systematic overvirial nature of cloud surveys, in 
several observational clump and core surveys, some objects appear to be 
{\it strongly} overvirial, exhibiting values of the {\it virial parameter}, 
$\alpha \equiv 5 \sigmav^2 R/GM \sim 10$--100 
\citep[see,e.g., fig.\ 16 of] [] {Barnes+11}, especially in the case of low-mass 
objects. These objects are traditionally interpreted as having 
a kinetic energy significantly larger than their gravitational energy, 
perhaps due to driving by stellar feedback, and therefore requiring 
confinement by external pressure to prevent them from dispersing,
as in the interpretation by \citet{Field2011} of the H09 sample. However, if 
we adopt the interpretation that star-forming GMCs are undergoing global 
and hierarchical collapse, pressure confinement is not satisfactory, 
since in this scenario the clumps should be gravitationally-dominated as 
well. Investigating the origin of these kinetic energy excesses within the 
scenario of collapsing clouds is one of the goals of this paper.

\subsection{This work} \label{sec:this_work}

In this work we create an ensemble of clumps in simulations of the
formation and evolution of molecular clouds, in order to investigate
their energy balance under a scenario of initial turbulence and
subsequent gravitational collapse. Our simulations, of course, have
a number of limitations, which are discussed in more detail in Sec.\
\ref{sec:caveats}, but here we note that they neglect magnetic
fields, stellar feedback, and have relatively low masses that restrict
the clumps and cores we obtain to values typical of low-mass star-forming
regions, which form low-mass stellar groups or low-mass clusters. 
Nevertheless, we expect that the results we obtain can be
extrapolated to regions of larger masses.

The organization of the paper is as follows. In Sec. \ref{sec:sim}
we briefly describe the simulations, and in Sec.\
\ref{subsec:tools}, we describe the clump-finding algorithm used to define 
clumps at various values of volume density, $\nth$, as well as the
selection criteria we used in order to avoid considering unrealistic
clumps (i.e. those that would be affected by stellar feedback in
reality). Next, in Sec.\ \ref{sec:res}, we present our results on the
energetics of the clumps and cores and their implications on Larson's
relations. In Sec.\ \ref{sec:discussion} we discuss our work in
context with recent related numerical studies, as well as the range of  
applicability, possible extrapolations,  and limitations of our study. Finally, 
in Sec.\ \ref{sec:concls} we present a summary and some 
conclusions.


\section{Numerical data and analysis} \label{sec:analysis}

\subsection{The simulations} \label{sec:sim}

The simulations used in this work are those presented in 
\citet{Gomez2014} and \citet{Heiner2014}. For historical reasons, in this 
paper we will refer to these simulations as \rv\ and \rt, respectively. Both
simulations were performed with the code Gadget-2 \citep{Springel2001}, 
using  296$^3 \approx$ 2.6 $\times 10^7$ SPH particles and a numerical 
box of 256 pc per side.
Each SPH particle is characterized by a single mass and
a smoothing length, the latter defined as the radius of the volume that 
contains $40\pm 5$ neighboring particles.
The simulations include the prescription for the formation of sink 
particles by \citet{Jappsen2005}, and the fix proposed by \citet{Abel11}, 
which eliminates several unphysical effects that arise in the standard SPH
prescription, and describes more accurately a number of physical 
instabilities, such as the Kelvin-Helmholz and Rayleigh-Taylor instabilities. 
Both simulations used the cooling and heating functions of \citet{KI02}, 
corrected for typographical errors as described in \citet{VS+07}.

In RUN20, the (uniform) initial density and temperature were set at 1$\pcc$ 
and 5206 K, and the mass per SPH particle is $\approx$ 0.02 $\Msun$ so 
that the total mass in the box is $5.26 \times 10^5 \Msun$. In this simulation, 
two cylindrical streams of warm neutral atomic gas, of diameter 64 pc and 
length 112 pc, were set to collide at the central ($x=0$) plane. In addition, 
a small amount of turbulent energy was added (with speeds $\thicksim 10 \%$ 
of the collision speed), at wave numbers $k=8$-$16\times 2 \pi/L$, where 
$L$ is the box size, so that the perturbations are applied at scales smaller than 
the inflow diameter. The collision produces a turbulent cloud, which grows in 
mass until it becomes gravitationally unstable and begins to collapse
\citep[e.g.,] [] {VS+07, HH08}. The collapse, however, is irregular and chaotic, 
because of the turbulence in the cloud, which creates a multi-scale and 
multi-site chaotic collapse \citep{Heitsch+08, VS+09}, rather than a monolithic 
one, producing a complex morphology, in which filamentary structures arise 
self-consistently. We refer the reader to \citet{Gomez2014} for more details.

On the other hand, \rt\ was produced with the aim of avoiding the
over-idealized conditions of a colliding-flow simulation, in which the
flows are perfect cylinders with circular cross-sections, moving in
opposite directions along the same axis. Instead, \rt\ was started by
applying a Fourier turbulence driver with purely solenoidal modes, with
wave numbers in the range $1<kL/2 \pi<4$ over the first 0.65 Myr of the
simulation, reaching a maximum velocity dispersion of $\sigma \approx$
18 km s$^{-1}$.  This produces a complex network of sheets and
filaments, which subsequently grow by gravitational accretion. In this
simulation, the initial density and temperature were set at 3 $\pcc$ and
730 K, respectively, the total mass in the box was $1.58 \times 10^6
\Msun$, and the mass per SPH particle was $\approx$ 0.06 $\Msun$.

Finally, in both simulations, the density threshold to form sink particles is set 
at $3.2 \times 10^6$ cm$^{-3}$, and no prescription for feedback is included. 
This implies that we must be careful in the choice of the clumps to be analyzed, 
in order to avoid including clumps that would have already been destroyed by 
stellar feedback, had it been included (see Sec.\ \ref{subsec:tools}). The main
parameters  of the simulations are summarized in Table \ref{tab:run_params}.

\begin{table*}[!lt]
   \begin{center}
	\begin{tabular}{| c | c | c | c | c | c | c |}
	\hline
	   	         &	$L$ [pc]	&	$T_{0}$[K]	&     $n_{0}$[cm$^{-3}$]	&     $m_{\mathrm{sph}}$ [$ \Msun$]  &  M$_{{\rm box}}$ [$\Msun$]  & Type		\\
         \hline
	RUN03	&	256	 	&	730			& 		3			&		0.06		&   1.5$\times 10^6$   &	Decaying turbulence		\\
	RUN20	&	256		&	 5206		&		1			&		0.02		&   5$\times 10^5$      &		Colliding flows		\\
	\hline
	\end{tabular}
   \end{center}
 \caption{ Initial conditions in simulations.  $L$ is the box size, 
		$m_{\rm sph}$ is the mass per SPH particle, $T_0$ is the
                  initial temperature, and $n_0$ is the initial density.}
  \label{tab:run_params}
\end{table*}

In these simulations, we analyze the physical properties of the clumps formed 
self-consistently, using a clump finding algorithm that we describe in 
Sec.\ \ref{subsec:tools}. We then measure the mass, size, density and 
velocity dispersion of the clumps in physical space
(not projected), to investigate 
their energy balance.  Table \ref{tab:run_stats_evol} gives the velocity dispersion, 
mass in sinks and global star formation efficiency, SFE 
$= M_*/(M_* + M_{\rm cold}$), of the simulations at the chosen times. Here, 
$M_{\rm cold}$ is the total mass in cold gas ($n > 50\, \pcc$, T$\sim$10-20 K).

\begin{table}[!lt]
  \begin{center}
  \resizebox{0.45\textwidth}{!}{
  \begin{tabular}{*{5}{|c}|}
  \hline
            & t (Myr) &  $\sigma_{v}$ (km/s) & M$_{sink}$ (M$_{\odot}$) & SFE \\
  \hline
            & 15.6    &  4.15   & 83.8  & 0.0002 \\
            & 18.2    &  3.69   & 512.9 & 0.0012 \\
 RUN03 	    & 18.5  &  3.65  & 576.9 & 0.0013 \\
            & 19.5  & 3.50 & 1300.0 & 0.0027 \\
            & 22.1 & 3.21 & 6786.9 & 0.1200 \\
\hline
            & 20.8 & 0.536 & 57.0 & 0.0032 \\
            & 21.1 & 0.533 & 119.6 & 0.0067 \\
RUN20       & 22.2 & 0.526 & 763.5 & 0.0409 \\
            & 24.8 & 0.518 & 2995.7 & 0.1471 \\
            & 26.5 & 0.517 & 4817.1 & 0.2234 \\
\hline
\end{tabular}
}
   \end{center}
\caption{Total velocity dispersion, mass in sinks, and total SFE ($= M_*/(M_*
+ M_{\rm cold}$)) at the times analyzed in the two simulations. Here,
$M_{\rm cold}$ is the total mass in cold gas ($n > 50\, \pcc$).}
\label{tab:run_stats_evol}
\end{table}

\subsection{Generation of the clump ensemble}
\label{subsec:tools}

In the present work, we use the term ``clump'' in a loose way,
simply to denote a local density enhancement above a given density
threshold, except for the objects defined at the highest thresholds
($\nth \ge 10^5\, \pcc$), which will be referred to as ``cores'', and
those defined at the lowest thresholds ($\nth = 300\, \pcc$), which will
be sometimes referred to as ``clouds''. No implication is made about a
specific density or mass range for the clumps, nor about whether they
will form only a few stars, or a cluster. This is a somewhat looser
usage of the term ``clump'' than the meaning adopted sometimes, of
clumps being the gaseous objects from which stellar clusters form
\citep[see, e.g., the review by] [] {BW99}. However, for the purposes of
the present study, which considers objects with a variety of densities
and masses, our more generic terminology is adequate. Note that, in
general, a single clump at a lower threshold may contain several clumps
at a higher one.

In what follows, we first describe the procedure for finding the clumps,
and then the selection criteria to include them in our sample.

\subsubsection{Clump finding algorithm}

The procedure to find clumps in the simulation was performed directly in the
SPH particles, without a previous mapping onto a grid. This allows the procedure 
to find the clumps in a manner independent of the grid resolution and without 
the smoothing inherent to the gridding procedure.

The procedure is the following.  First, we select all the SPH particles
in the simulation with density above a certain threshold, $n_{th}$. We
then locate the particle with the highest density.  This particle and
all those located within its smoothing length are labeled as
members of the clump. Then, the following steps are iterated: we locate
the member of the clump with the highest density to which this
subprocedure has not been applied, and then label as members all the
particles within a smoothing length not-yet belonging to the clump. The
iteration ends when all the clump members are examined. If there are
particles remaining with density $n > n_{th}$ that are not yet
members of any clump, we locate the one with the highest density and use
it to define a new clump and the procedure is repeated.

In summary, this procedure finds the largest connected object above a volume 
density threshold.

\subsubsection{Clump sample} \label{sec:clump_sample}

For the analysis of both simulations we only considered clumps with more 
than 80 SPH particles ($ M \ge$ 4.8 $\Msun$ in RUN03 and $ M \ge$ 1.6 
$\Msun$ in RUN20), in order to guarantee that they are well resolved, 
according to the criterion by \citet{BB97}. This amounts to twice the number 
of particles contained within one particle smoothing length.

Since in \rt\ we first apply a turbulent driver, and then we leave the 
simulation to decay and collapse, we analyze various timesteps that correspond 
to different levels of the turbulence and different evolutionary stages of the 
clouds and clump populations. Specifically, we analyze the simulation at 
$t=15.6$, 18.1, 18.5, 19.5,  and 22.1 Myr.  Sink (``star'') formation begins in the 
simulation at $t \sim 14.7$ Myr, and at $t = 22.1$ the total mass in sinks is $6787 \Msun$, 
amounting to $\sim 0.4\%$ of the total mass in the box, and $\sim 4\%$ of the 
cold gas mass. At these times, we generate an ensemble of clouds, clumps and 
cores by applying the clump-finding algorithm at  the density thresholds,  
$n_{\rm{th}}= 300$, $10^3$, $3 \times 10^3$, $10^4$ and $10^5\, \pcc$,  
for both simulations. 

In RUN20 the turbulence in the cloud develops self-consistently, and so it never 
is excessive (i.e., not larger than what would be produced self-consistently by 
the gravitational contraction). Nevertheless, the structures (filaments and 
clumps) still evolve, the filaments gathering mass by accretion from their 
surrounding medium, and the filaments themselves feeding the clumps inside 
them \citep{Gomez2014}. 
However,  close inspection of the evolution of the filament/clump systems shows 
that they form at about the same time, and the flow along the filaments toward 
the clumps develops later \citep[see also] [] {GO15}. This will be relevant 
later in the investigation of  whether the accretion from the filaments onto the 
clumps has any effects on the dynamics of the latter (Sec. \ref{subsec:filament}). 
In this simulation we choose timesteps that exhibit well-defined filament/clump 
systems, in order to study whether clumps accreting from filaments exhibit 
systematically larger velocity dispersions than those expected from energy
equipartition considering the mass of the clump only. Specifically,  we consider 
time steps at $t=$ 20.8, 21.2, 23.2, 24.8 and 26.5 Myr.

In order to find cores of high column density we selected some timesteps
\footnote{It is important to note that the original timestep between successive 
dumps in both simulations was $\Delta t =0.136$ Myr. This timestep proved
inadequate to resolve the formation and collapse of the densest 
clumps which, at $n > 10^6\, \pcc$, have free-fall times $\tff \lesssim 0.045$ 
Myr.  Therefore, to find cores of high column density we restarted the 
simulations shortly before new sinks appeared  with 10 times finer temporal
resolution.}, $t=$ 18.1 and 18.5 Myr for \rt, and $t=$ 20.8 and 21.2 Myr for \rv,  
and we applied the clump-finding algorithm at $n_{\mathrm{th}}=$3x$10^5$ 
and $10^6$.

In the SPH simulations used in this work, the SPH particle mass ($m_{\rm SPH}$; 
see Table \ref{tab:run_params}) is fixed. Thus, for clumps satisfying the above 
conditions, we compute the gas mass as $M_{\rm gas} = \Ncl m_{\rm SPH}$, 
where $\Ncl$ is the number of SPH particles contained in the clump.
Note that , in general, the measured masses are smaller at higher thresholds, 
because denser objects are embedded within larger ones that  are more massive, 
but less dense on average. 

\begin{figure*}[!lt]
\centering
  \includegraphics[width=0.8\textwidth]{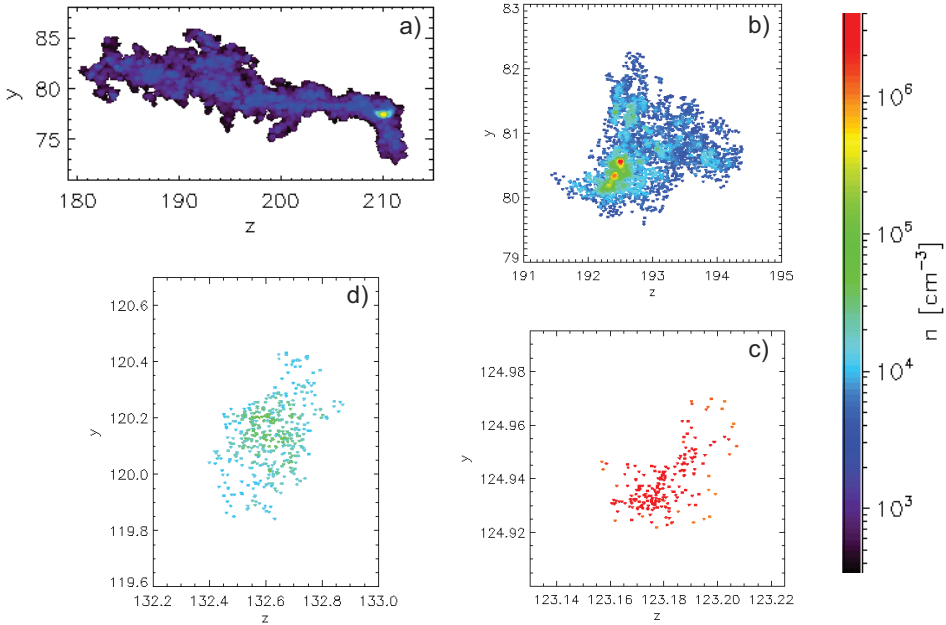}
   \caption{Clumps at different density thresholds and snapshots in both simulations
                 at the same density scale. a) Clump in RUN03 at $n_{th}=300$  $n_{0}$; b)
                 RUN20, $n_{th}=10^3$  $n_{0}$; c) RUN03, $n_{th}=3\times 10^3$ $n_{0}$;
                 d) RUN20 $n_{th}=10^4$ $n_{0}$. We find, in general, more elongated 
                 structures than spherical.}
  \label{fig:clumps}
\end{figure*}

Finally, note also that the clumps exhibit complex morphologies, being
far from spherical in general, and often display elongated and twisting
shapes, as shown in Fig. \ref{fig:clumps}. This implies that there is an
inherent ambiguity in the definition of the clump size, since in general 
they have more than one characteristic dimension.  With this caveat in 
mind, we compute the clump ``radius'' as $R= (3V/4\pi)^{1/3}$, where 
$V$ is the sum of the specific volumes of all particles that belong to a
clump, i.e.,

\begin{equation}
V =  \sum_{i=1}^{\Ncl}V_{i} = \sum_{i=1}^{\Ncl}\frac{m_{\rm sph}}{\rho_{i}}= 
                                               m_{\rm sph}\sum_{i=1}^{\Ncl}\rho_{i}^{-1},
\end{equation} 
where $\rho_{i}$ is the density of
particle $i$. 
With this definition, the mean density of the clump ($\bar{\rho} = M/V$) is given by,
\begin{equation}
\bar{\rho} = \frac{ m_{\rm sph}\Ncl }{ m_{\rm sph} \sum_{i=1}^{\Ncl}\rho_{i}^{-1}}= 
\frac{\Ncl}{\sum_{i=1}^{\Ncl}\rho_{i}^{-1} }.
\end{equation}

Since feedback is not included in either of our simulations, we need
to apply some kind of criterion to avoid including clumps that exhibit
unrealistic physical properties because feedback should be already
dominating their dynamics if it had been included. We therefore consider
only clumps that, at a density threshold of $\nth = 10^5\, \pcc$, have
a star formation efficiency SFE$_5 < 65\%$, where the SFE$_5$ is defined as
\begin{equation}
{\rm SFE}_5 = \frac{M_*} {M_{{\rm tot},5}},
\label{eq:SFE}
\end{equation}
and $M_{\rm tot} = M_* + M_{{\rm gas},5}$, $M_*$ is the mass in stars (sink
particles) and $M_{{\rm gas},5}$ is the mass in dense gas above $n
  = 10^5\, \pcc$. We choose this
value as a compromise between realistic SFEs for massive cluster-forming
clumps \citep[10--50\%, ][]{Matzner00, LL03} and obtaining a reasonably 
large statistical clump sample.

For lower clump-defining density thresholds, we must impose a
further restriction on the maximum accepted SFE at each threshold. 
Star formation is a highly spatially intermittent phenomenon,
so that star-forming sites only occur at a few and highly localized
positions that have the highest densities in a large MC. Thus, if one
focuses on a given star-forming site and measures the gas mass around it
at various thresholds, this mass will be larger for lower thresholds,
since this procedure includes progressively more material from
progressively larger distances from the star-forming site. In
consequence, the measured SFE around a star-forming site will be smaller
at lower thresholds (as long as no other site enters the domain defined by the
threshold).  This is consistent with the general trend that lower-density
objects generally are observed to have lower SFEs \citep[e.g.,] []
{Palau+13, Louvet+14}.

\begin{figure*}[!lt]
\centering
  \includegraphics[width=0.9\textwidth]{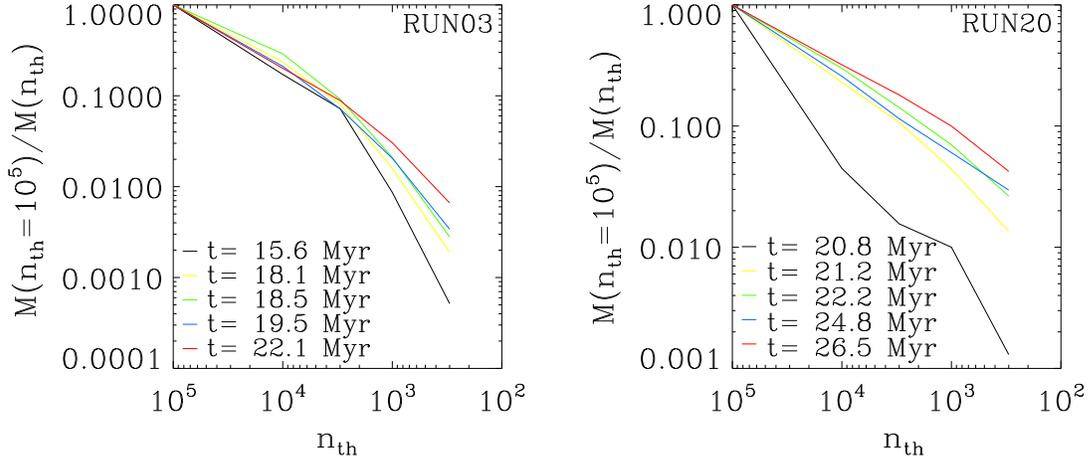}
  \caption{ Maximum SFE (SFE$_{max,i}$) allowed for clumps included in 
      the sample
      as a function of density threshold, $\nth$, 
      as given by eq.\ (\ref{eq:SFE_weight}), at the various
      timesteps considered in the two simulations. }
  \label{fig:massratio}
\end{figure*}

To replicate this trend, we require progressively smaller efficiencies
at lower thresholds in order to admit a clump in our sample. Thus,
the maximum star formation efficiency, SFE$_{{\rm max},i}$, 
for any clump included  
in our sample at threshold $i$ is given by
\begin{equation}
\hbox{SFE}_{{\rm max},i} =   \left(\frac{M_5}{M_i} \right) \hbox{SFE}_{5}, 
\label{eq:SFE_weight}
\end{equation}
where $M_5$ is the total mass in clumps defined at threshold $\nth =
10^5\, \pcc$, $M_{i}$ the total mass in clumps defined at the $i$-th
threshold, $n_{{\rm th},i}$, and SFE$_{5}=65\%$.  Figure
\ref{fig:massratio} shows the mass fraction $M_5/M_i$ as a function of
$n_{{\rm th},i}$ for the two simulations. With this prescription,
we avoid including clumps whose measured SFEs at low thresholds are
so large that at a higher threshold they would exceed the maximum SFE
allowed for it.

\section{Results}
\label{sec:res}

\subsection{Testing for Larson's relations}
\label{subsec:larson_fall}

\subsubsection{The density-size relation} \label{sec:rho-R}

\begin{figure*}[!t]
\centering
         \includegraphics[width=0.85\textwidth]{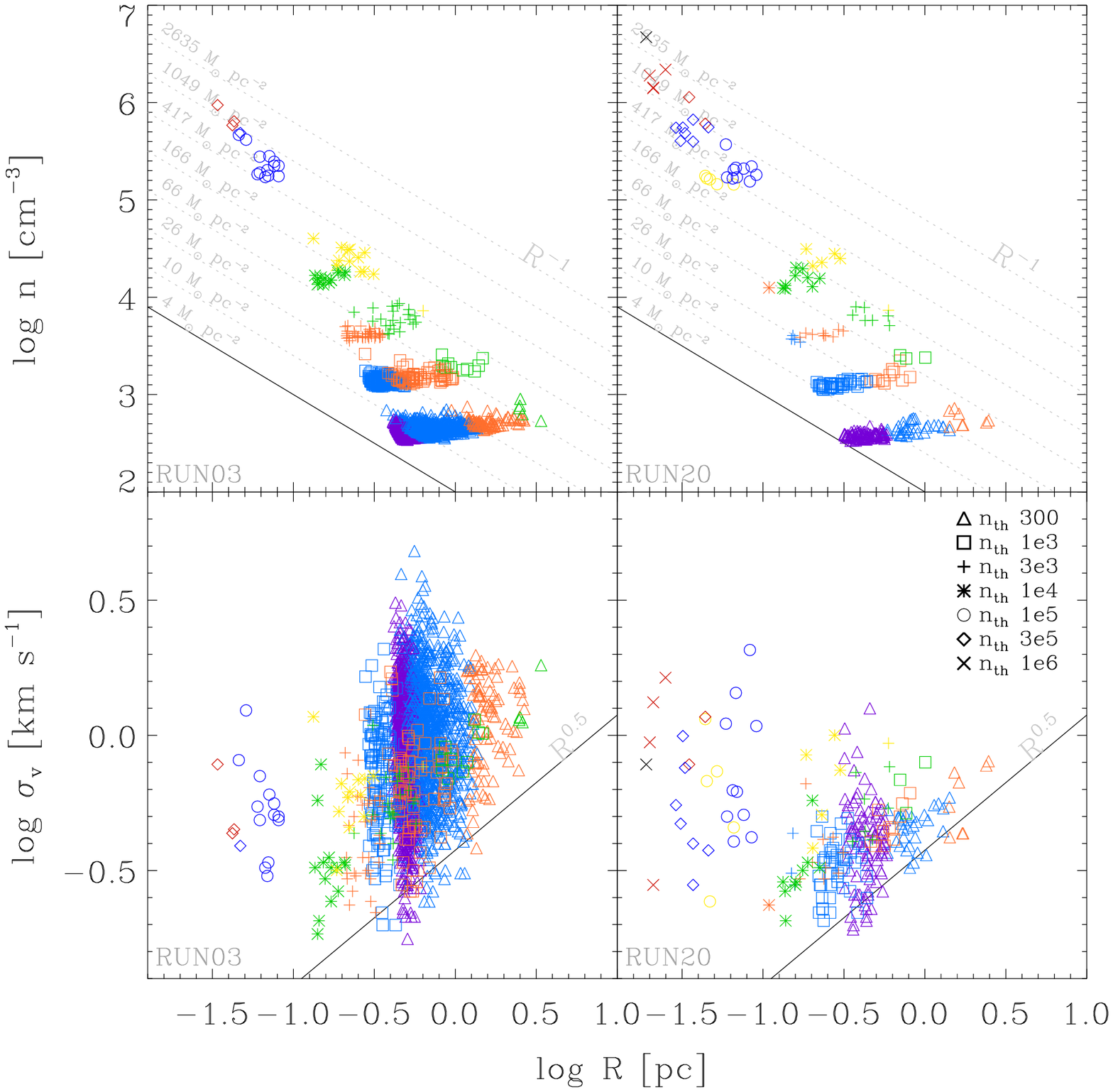}
         \caption{ Larson-like relations for simulated clumps and cores. 
         Symbols correspond to the volume density threshold used to 
         define clumps. 
         The dotted lines correspond to constant column density values and 
         colors represent the different column density-ranges,
         bounded by the values of the
         dotted lines. Purple:
         $\Sigma < 10 \Msun $pc$^{-2} $; blue: $10 < \Sigma < 26  \Msun $pc$^{-2} $;
         orange:  $26 < \Sigma < 66 \Msun $pc$^{-2}$; green: $66 < \Sigma < 166  \Msun $pc$^{-2} $;
         yellow: $166 < \Sigma < 417  \Msun $pc$^{-2} $; blue:  $417 < \Sigma < 1049  \Msun $pc$^{-2} $;
         red:  $1049 < \Sigma < 2635  \Msun $pc$^{-2} $, and black:  $\Sigma > 2635  \Msun $pc$^{-2} $. 
         Note that, by selecting clumps by column density, each sample follows
          a Larson-like density-size relation.} 
         \label{fig:Larson}
\end{figure*}

We first check whether our clump sample, occurring in clouds undergoing 
global collapse, satisfies Larson-like relations. Figure \ref{fig:Larson} shows 
our clumps in the $n$ {\it vs.} $L$ and $\sigmav$ {\it vs.} $L$ diagrams for 
the two simulations.  In this plot, the different colors represent different
column densities, the dashed lines represent various column densities, and 
the symbols correspond to the volume density thresholds used to define 
the clumps.

Due to the resolution requirement that our clumps contain at least 80 
SPH particles, our sample is mass-limited from below, so that the clumps 
are constrained to have masses $M_{\rm cl} \ge 1.6\, \Msun$ in RUN20 and
$M_{\rm cl} \ge 4.8\, \Msun$ in RUN03. 

A first point to notice in Fig. \ref{fig:Larson} is that RUN03 produces more 
clumps than RUN20.  This is because in RUN03 the density structures are 
scattered throughout the simulation domain, since the clouds and clumps 
have been produced from a large 
number of turbulent fluctuations in the early stages of the run. Instead, in 
RUN20 there  is only a single large cloud complex, produced initially by the 
collision of the large cylindrical WNM streams. 

From Fig. \ref{fig:Larson}, we also see that our clump sample does not follow 
the standard Larson $n \propto R^{-1}$ relation (Fig.\ \ref{fig:Larson}). 
Instead, the entire sample occupies a triangular region in the $n$-$R$ diagram, 
so that no single density-size relation exists. Additionally,  we observe a very 
well defined group of points at nearly constant density for each volume density 
threshold. However, we also observe that, when the clumps are classified by 
column density (with various column density ranges shown as the different 
colors in the figure), then each sub-sample traces a slope $R^{-1}$ as in the
\citet{Larson1981}  density-size relation. This supports the notion that 
the density-size relation is an artifact of defining clumps by a column density 
threshold.

The fact that the clumps defined at a certain volume density threshold appear 
to have a nearly constant {\it volume} density was already noticed in numerical 
simulations by \citet{VBR97} and \citet{Ballesteros2002}, and later interpreted 
by \citet{Ballesteros2012} and \citet{Beaumont+12} as a consequence of the
steep slope of the high-density side of the density PDF, which implies
that most of the volume (and even the mass) is at the lowest densities
allowed by the threshold, if the threshold is above the density corresponding 
to the maximum of the PDF. The same is expected to happen for the column 
density, if it is also described by a lognormal or a power law with a slope 
steeper than $-1$ \citep{Ballesteros2012}, again reinforcing the notion that 
the apparent constant column density of molecular clouds is an artifact of the
restricted column density range allowed by the tracers used to observe them, 
such as the $^{12}$CO line.\footnote{ It has been argued by \citet{Lombardi+10} 
that the constant column density of molecular clouds is an actual physical
property of molecular clouds, which can be observed in dust extinction maps 
that allow thresholds significantly lower than that claimed for the physical 
column density of the molecular clouds. However, the result by 
\citet{Ballesteros2012} that the observed most common column density for a 
cloud is dominated by the lowest densities present in the cloud when the PDF 
slope is $< -1$ and by the highest densities otherwise can explain the result by
\citet{Lombardi+10}, if the bimodal pdf of the Galactic gas is taken into account 
\citep[e.g.,] [] {VS+00b, AH05, Gazol+05}. We plan to address this issue in a 
separate contribution.}

\begin{figure*}[!t]
\centering
         \includegraphics[width=0.85\textwidth]{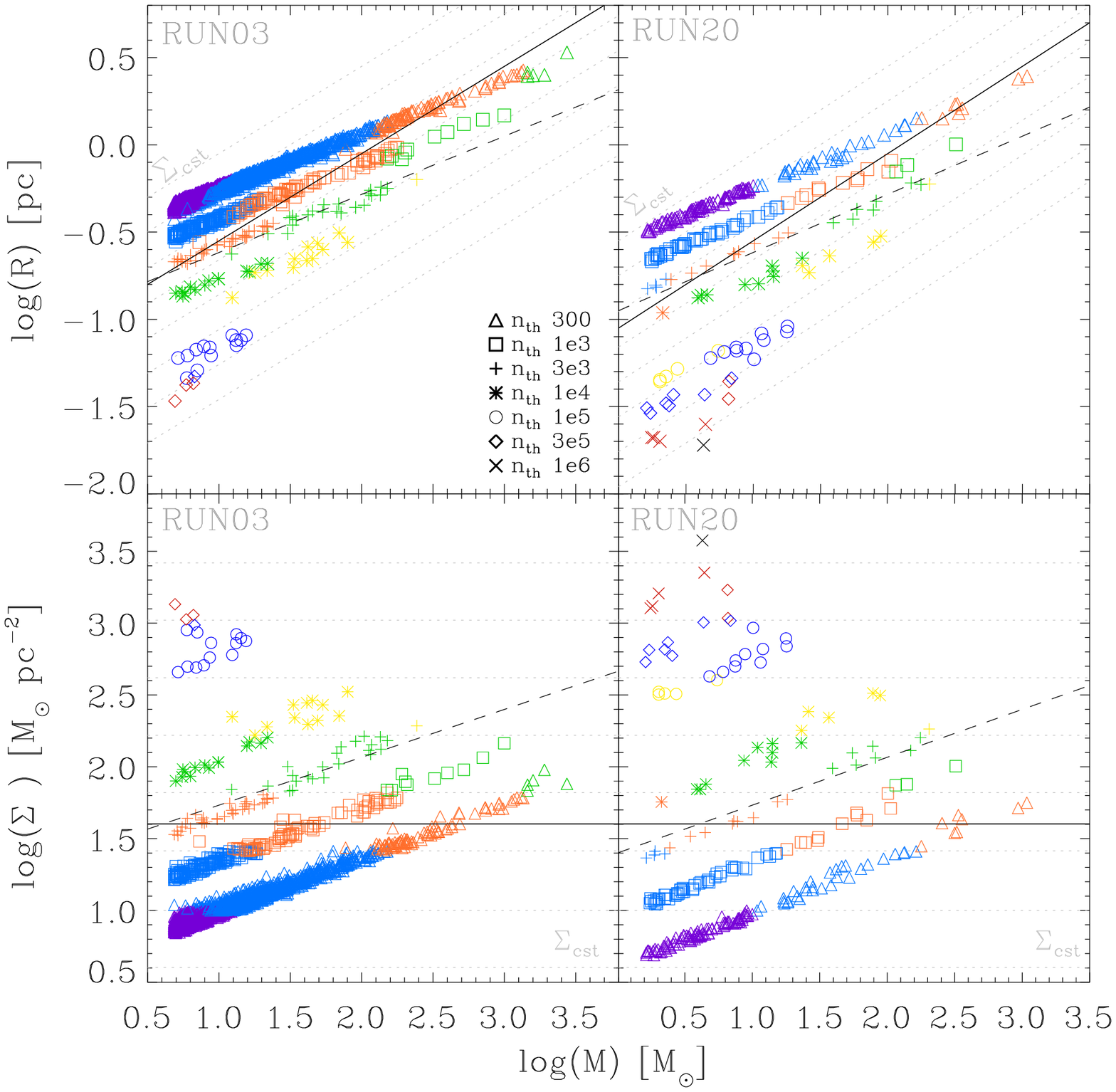}
       
         \caption{ Size and column density 
            {\it vs.} mass. The solid lines show the scaling
             implied by the density-size
               Larson-like relation; the dashed lines show the
               slope 1/3 corresponding to constant volume  density.
               The color
               scheme, the symbols and the values of
             $\Sigma$ (dashed gray
             lines) are the same used in Fig. \ref{fig:Larson}.  Note
             that the clumps belonging
               to a given $\Sigma$ range
             follow a scaling similar to that implied by Larson's
             density-size relation. }
      \label{fig:Fall}
\end{figure*}

An expression equivalent to Larson's density-size is that of mass {\it vs.}  
size, which reads $M \propto R^2$. Figure \ref{fig:Fall} shows the scaling of 
size {\it vs.} mass (left panels) and also the column density {\it vs.} mass (right 
panel) for our clump sample in the same $\Sigma$ ranges as in Fig. 
\ref{fig:Larson}.  Solid lines correspond to $M \propto R^2$ 
(or constant-$\Sigma$). Again, it can be seen that, in both simulations, when the 
entire sample is considered, the clumps do not show constant column density 
(right panels, Fig.\ \ref{fig:Fall}) nor an $M \propto R^2$ scaling (left panels, 
Fig.\ref{fig:Fall}). However,  such a correlation reappears when the data are classified by column density.
On the other hand, at every volume density threshold (different symbols for 
each threshold) clumps show the relation $R\propto M^{1/3}$, denoted
by the dashed black line in
Fig. \ref{fig:Fall}, which is the scaling expected for clumps of constant 
volume density \citep[][]{Ballesteros2012, Beaumont+12}.

It is worth noting in the {\it right} panels of Fig. \ref{fig:Fall} that
clumps in a given range of column densities include clumps defined at
various {\it volume} density thresholds. Nevertheless, there is a net
trend for clumps defined at higher volume density thresholds to fall in
higher column density ranges. This behavior is also present in
observational data \citep[see, e.g., Figure 14 of] [] {Barnes+11}.

\subsubsection{The linewidth-size relation} \label{sec:sigma-R}

With respect to the $\sigmav$-$R$ relation, we notice in Fig.\
\ref{fig:Larson} that the $\sigmav \propto R^{1/2}$ scaling is not
satisfied by the whole clump sample in either of the simulations and,
instead, the ensemble of clumps fills a large area in the $\sigmav$-$R$
diagram \citep[as is often the case in observational surveys as well; see,
  e.g., ] [and references therein]{Heyer+01, Ballesteros2011, Heyer15}.
However, it can be observed that the clumps with $\log(R/{\rm
pc}) \gtrsim 0.5$ are bounded from below by approximately the Larson
slope (right panels of Fig.\ \ref{fig:Larson}). A similar effect was
observed by \citet{VBR97}. In addition, it is also noticeable that some
of the samples at certain $\Sigma$ ranges (see, for example the orange
and green points) seem to follow this scaling. In Sec.\
(\ref{sec:scatter}) we discuss the origin of the observed scatter.

\subsection{Generalization of Larson's relations and energy balance}
\label{subsec:heyer_alpha}

\subsubsection{The KH diagram} \label{sec:KH_diag}

The results from the previous section show that the clumps defined
by volume density thresholds in the clouds in our simulations of global,
hierarchical gravitational collapse do not seem to follow the
linewidth-size relation. Instead, Larson-like density-size (and
  equivalent) relations appear when
  the selection of the clumps is done by means of a {\it column} density
threshold or range. Larson-Like linewidth-size relations appear for {\it
some} of the clumps. 

We now test whether our clump sample follows a relation of the form of
eqs.\ (\ref{eq:gener_Larson}) or (\ref{eq:free-fall}); i.e., we test for
whether the clouds appear to be near virial equilibrium or energy
equipartition, respectively, where the latter is consistent with
free-fall. These relations can be considered as the generalization of
Larson's relations when the column density of the objects in the sample
is not constant.

\begin{figure*}[!lt]
\centering
         \includegraphics[width=0.95\textwidth]{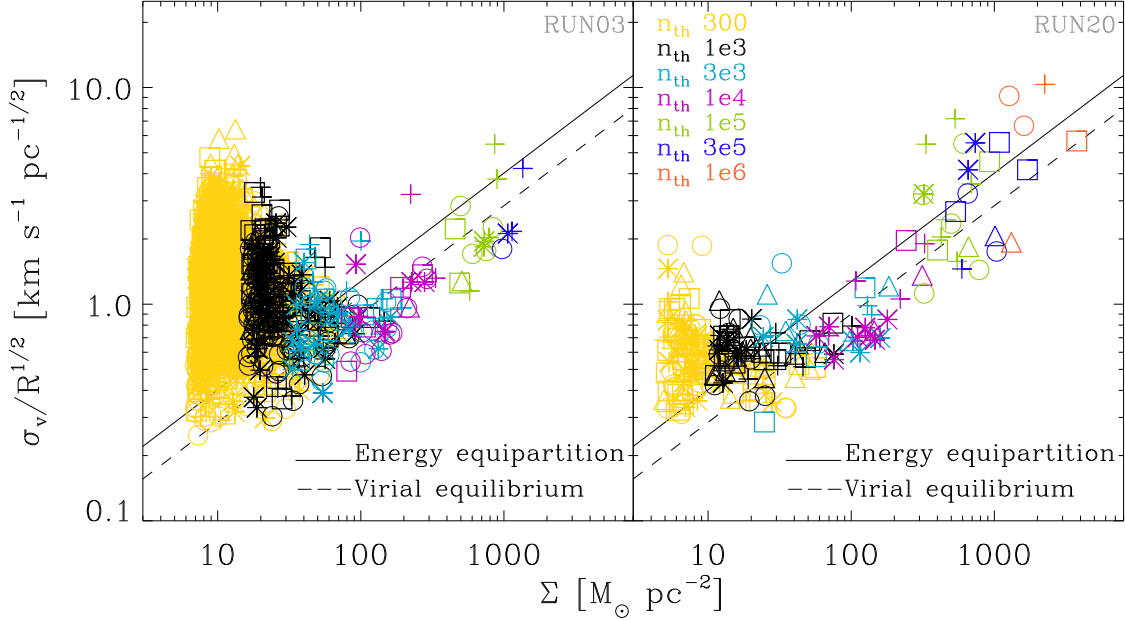}
         \caption{Generalization of Larson's scaling
           relations. The clouds at $n_{\rm th}=300n_{0}$
           can be considered as atomic
           because of their low column
           density. Here, colors represent the different volume density thresholds
           used to define clumps with the clump-find algorithm described in section
           \ref{subsec:tools}. Symbols represent different time-steps for RUN03 ( $\triangle$=15.6,
           $\square$=18.1,$+$=18.5, $\ast$=19.5, $\bigcirc$=22.1) and RUN20 
           ( $\triangle$=20.8, $\square$=21.2,$+$=22.2, $\ast$=24.8, $\bigcirc$=26.5). }
      \label{fig:Heyer}
\end{figure*}

Figure \ref{fig:Heyer} shows the ratio $\sigma/R^{1/2}$ {\it vs.} $\Sigma$
for clumps in \rt\ and \rv.  In what follows, we refer to this diagram as the 
Keto-Heyer (or KH) diagram \citep[] [H09] {KM86}. The solid line
corresponds to virial equilibrium and the dashed line corresponds to energy 
equipartition, or free-fall condition. 
Contrary to figures \ref{fig:Larson} and \ref{fig:Fall}, in these plots we use  
colors to denote the volume density thresholds at which clumps were defined, 
and different symbols to represent the timesteps considered in each simulation.
Column density has been computed as $\Sigma = M_{g}/\pi R^2$  where $M_{g}$ 
represents the gas mass in the clump (we will consider the stellar mass---i.e. sink 
mass, $M_{*}$---later).

\subsubsection{Low-column density clumps} \label{sec:low_col_dens}

The first noticeable feature in these plots is the group of clumps at
the lowest volume density threshold, i.e. $n_{\rm th} = 300 n_{0}$ (yellow
symbols), which also have the lowest surface
densities.  These clumps are seen to occupy a region in the diagram that
extends from the virial 
and free-fall lines to over one order of magnitude in $\sigmav/R^{1/2}$
above those lines, at roughly constant column density. We note that this
is precisely the kind of behavior displayed in observational clump
surveys (see, e.g., Figure 10 of \citet{KM86} and
Figure 13 of \citet{Leroy+15}).

\begin{figure*}[!lt]
\centering
         \includegraphics[width=0.95\textwidth]{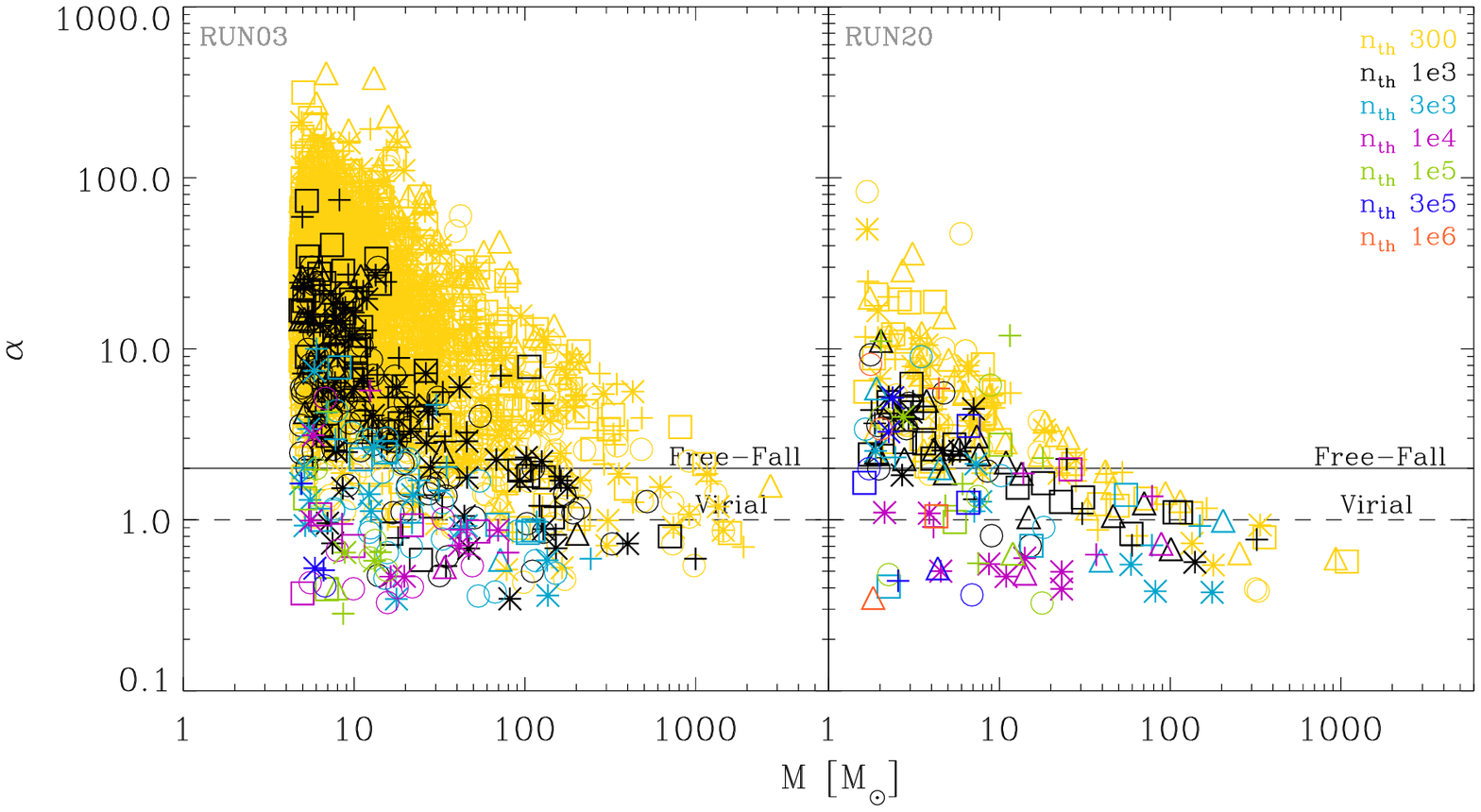}
     \caption{$\alpha$ parameter as a function of mass for the clumps for
       both simulations. Symbols represents different time steps, RUN03 ( $\triangle$=15.6,
           $\square$=18.1,$+$=18.5, $\ast$=19.5, $\bigcirc$=22.1) and RUN20 
           ( $\triangle$=20.8, $\square$=21.2,$+$=22.2, $\ast$=24.8, $\bigcirc$=26.5). }
      \label{fig:Alpha}
\end{figure*}

Another frequent way of displaying the energy balance of the clumps is
by plotting the ``virial parameter'' $\alpha = 2 \Ek/|\Eg| = 5 \sigmav
R/GM$ \citep{BM92}. Figure \ref{fig:Alpha} shows this parameter versus
the clump mass for the two simulations. As in Fig. \ref{fig:Larson}, the clumps with
the lowest masses exhibit the largest scatter in $\alpha$, with excesses
of up to nearly two orders of magnitude. Again, this is similar to the
observed behavior of clump surveys \citep[see, e.g., figure 16 of] []
{Barnes+11}.

This behavior is generally interpreted as implying that these clouds
have an excess of kinetic energy over their gravitational energy,
therefore being unbound, and needing to be confined by an external
pressure to avoid rapid dispersal or else that they are being
dispersed by the energizing action of stellar feedback. This latter
possibility is not possible in our simulations, because we have not
included any kind of stellar feedback, with the purpose of examining
the action of only the initial assembling turbulent motions and of the
gravitationally-driven motions.

An alternative possibility is that, if turbulent compressions in the
atomic interstellar medium (ISM) are causing the early assembly stages
of these clouds, their associated velocities may be larger than the
corresponding gravitational velocities for those objects, although in
this case their role is to {\it assemble} the clouds rather than to
disperse them, in the context of {\it converging flows} from large-scale
turbulent fluctuations. Later, as a cloud gains mass, its gravitational
velocity may begin to dominate over the initial turbulent compression
that started it, which may itself tend to dissipate.

\begin{figure*}[!t]
\centering
\includegraphics[width=0.45\textwidth]{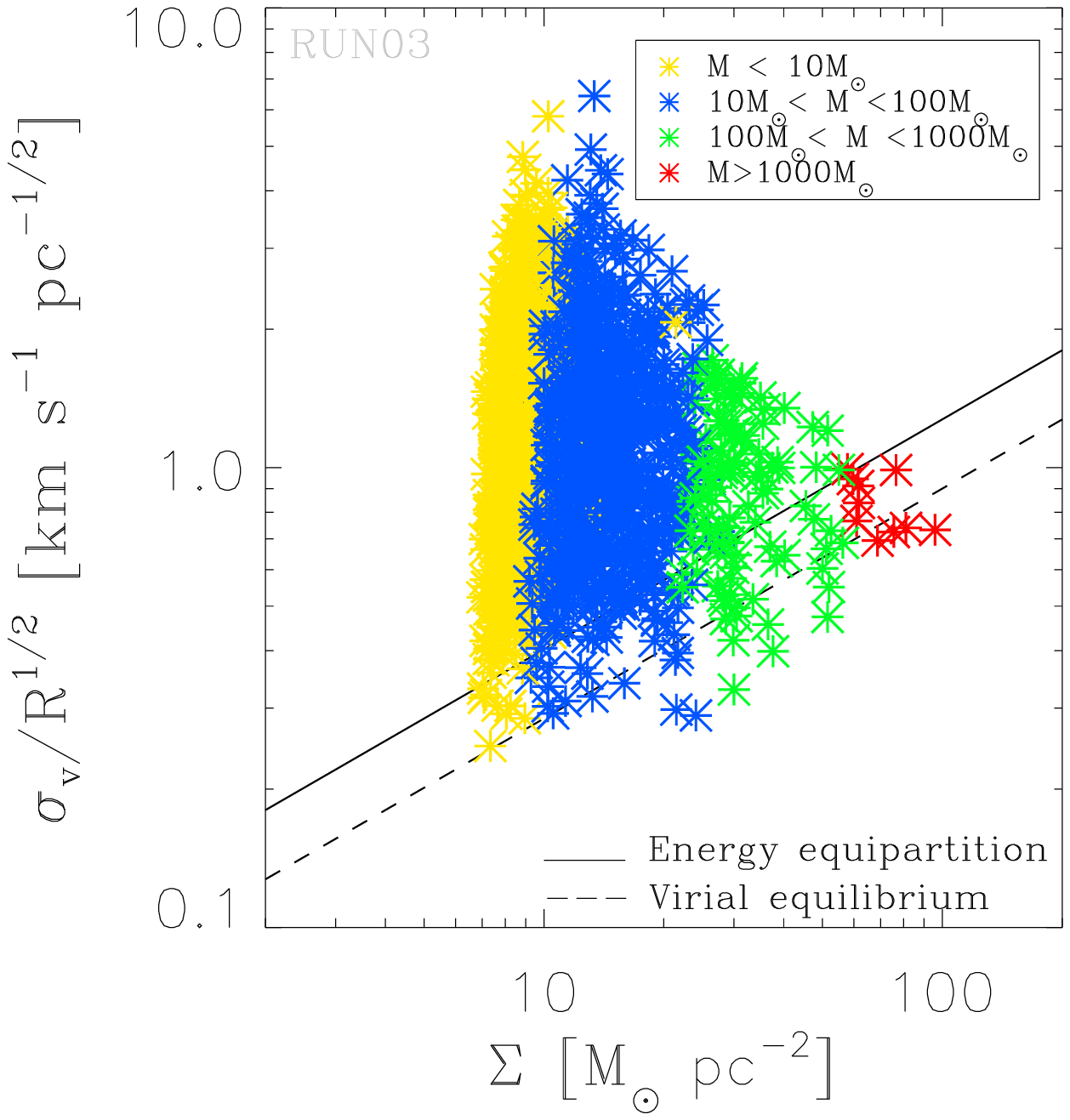}
\includegraphics[width=0.45\textwidth]{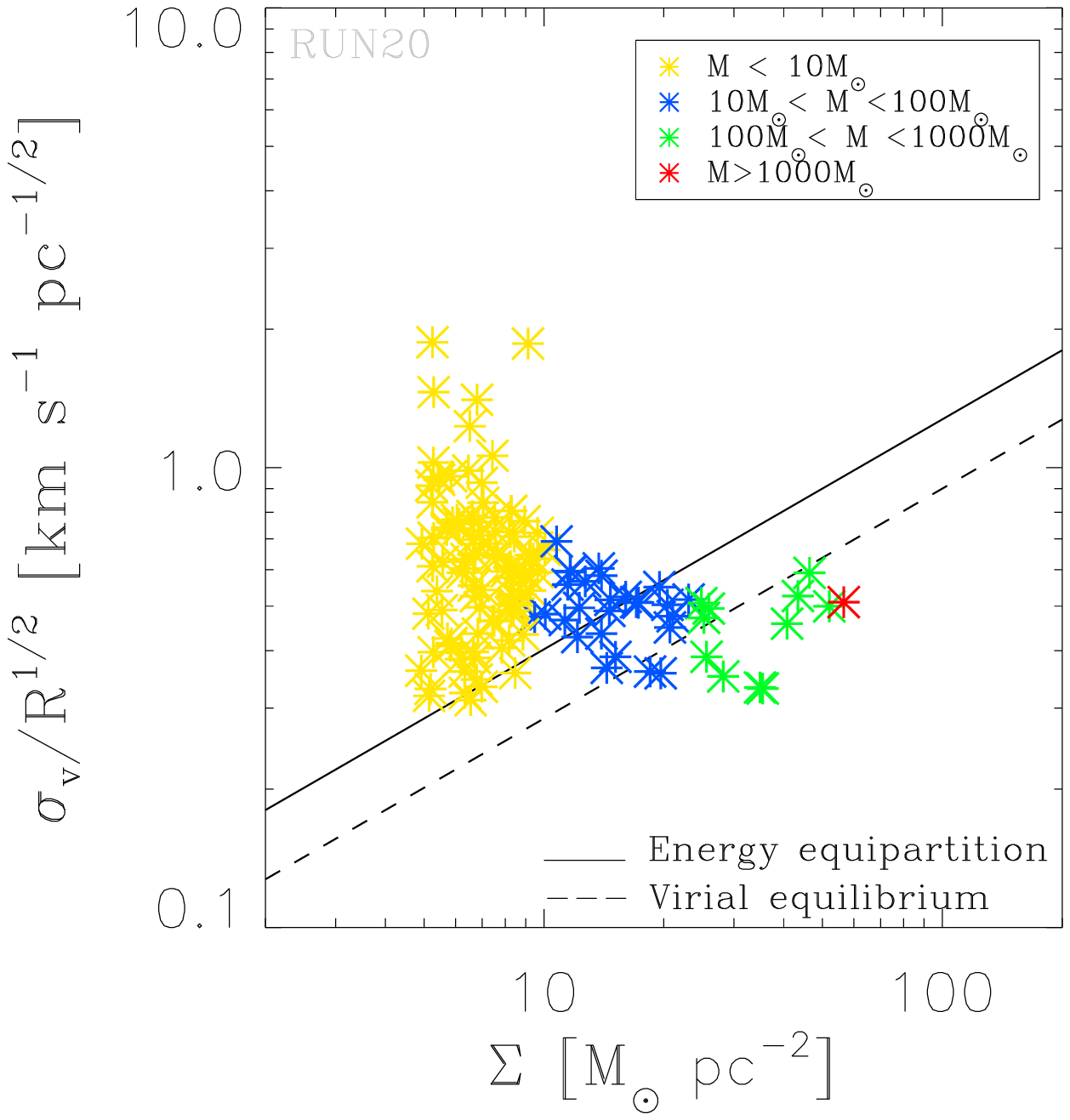}
\includegraphics[width=0.42\textwidth]{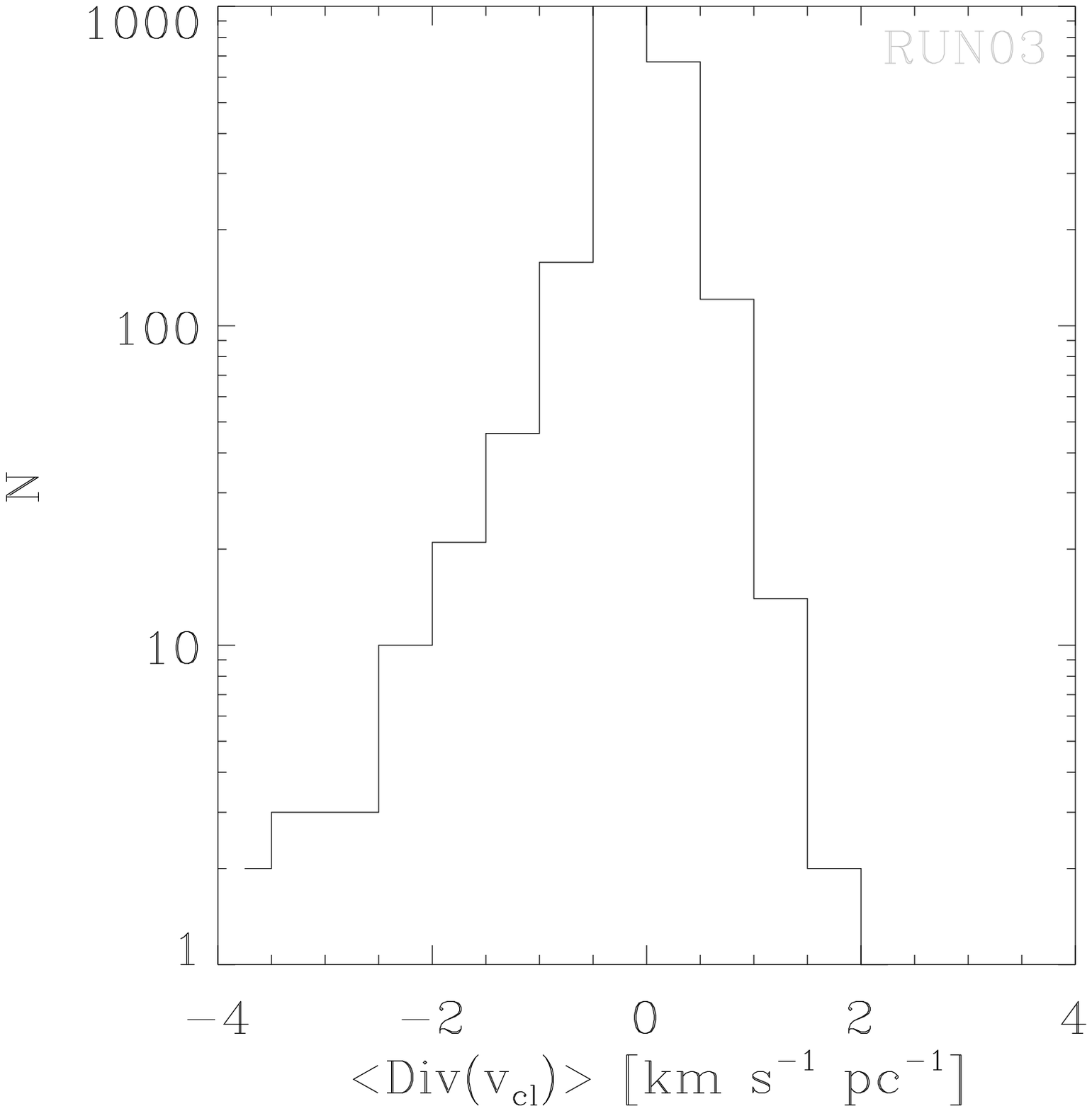}
\includegraphics[width=0.42\textwidth]{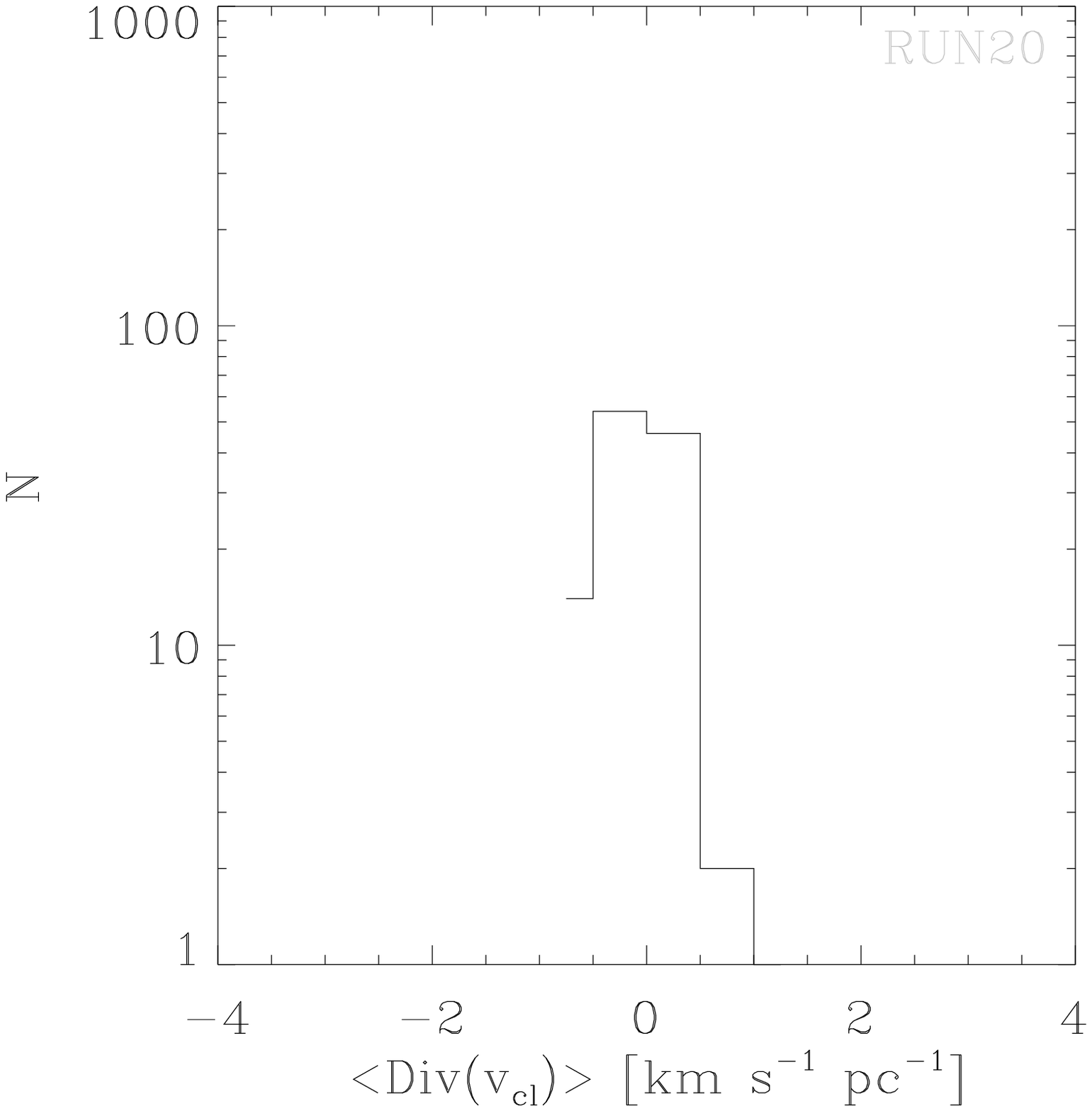}
\caption{{\it Top:} Clumps defined at the lowest density threshold 
        ($\nth = 300\pcc$) plotted in the KH diagram.  
        Clumps at the lowest volume density threshold generally have the 
        lowest column densities as well.
        Different colors represent clumps in different mass ranges. The most
        massive clumps are seen to lie closest to the virial and equipartition
        lines. The {\it bottom} row shows the histograms of the mean velocity
        divergence in the clumps defined at $\nth = 300\, \pcc$ and with
        masses  $M < 100 \Msun$, in \rt\ ({\it left}) and in RUN20 ({\it
        right}).}
\label{fig:KH_low_nth}
\end{figure*}

To test for this, in the {\it top row} of Fig.\ \ref{fig:KH_low_nth} we
show the clouds obtained at the lowest density threshold ($\nth = 300\,
\pcc$) in the KH diagram, but representing different mass ranges with
different colors. Note that colors in these plots are also representative of 
sizes ranges, given that all these clumps have similar volume densities (cf.\ 
Sec. \ref{sec:rho-R}), and therefore the most  massive ones are also the 
largest ones. The most massive clumps are seen to lie closest to the virial
and equipartition lines, suggesting that these objects tend to be dominated 
by the gravitational velocity rather than by the turbulent velocity. This 
means that they have the largest gravitational velocities at a given column 
density, consistent with an evolutionary picture where the clumps are first 
assembled by large-scale turbulent compressions and, as they grow, they 
change from turbulent assembly to gravitational contraction.

This scenario can be further tested by measuring the mean velocity
divergence in each clump \footnote{The particle velocity divergence 
was obtained directly from the GADGET-2 code, which computes the 
divergence in terms of the kernel function, the density and the velocity of 
each particle. Therefore, the errors in the velocity divergence are of the 
same order as those in the integration of the equations.}, 
which shows  whether the clump is contracting or expanding as a whole.
A negative mean divergence means that the clump is contracting on average 
and, if its velocity is not driven by gravity, then its contraction must be a 
turbulent compression from the outside gas \citep[e.g.,] [] {VS+08, GS+14, Pan+15}. 
The {\it bottom row} of Fig.\ \ref{fig:KH_low_nth} shows the histograms of 
the mean velocity divergence of all the clouds defined at the lowest 
threshold ($\nth = 300\, \pcc$) and with masses $M < 100 \Msun$ which 
are the clumps exhibiting the largest scatter in the KH diagram. We see that 
more than half of the clumps ($\sim 60 \% $ in both simulations) 
have negative divergence, indicating that they are contracting on 
average, and therefore are in the process of {\it assembly}, although 
significant fraction is undeniably in the process of {\it dispersal}. 

\begin{figure*}[!t]
\centering
\includegraphics[width=0.4\textwidth]{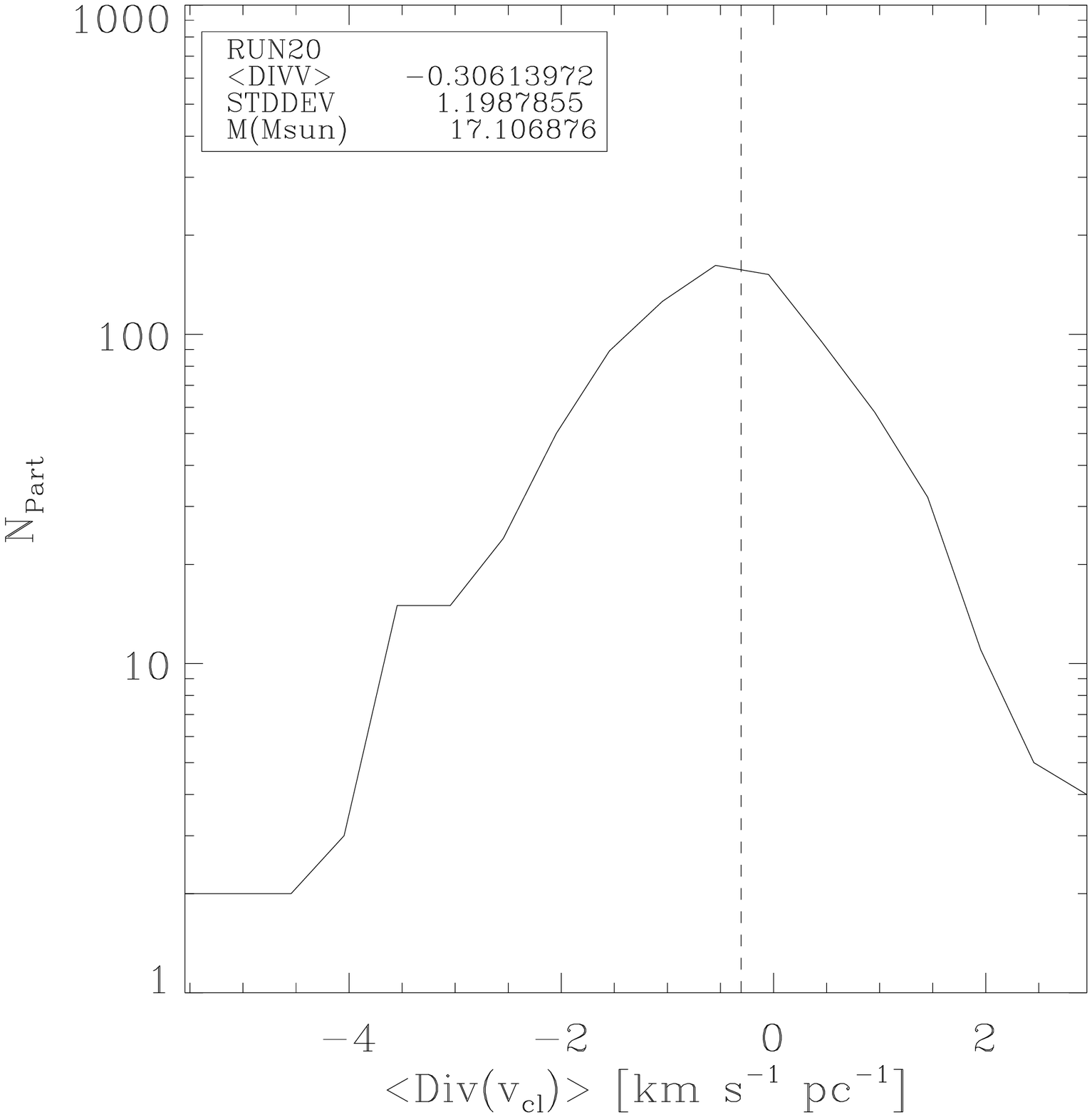}
\includegraphics[width=0.4\textwidth]{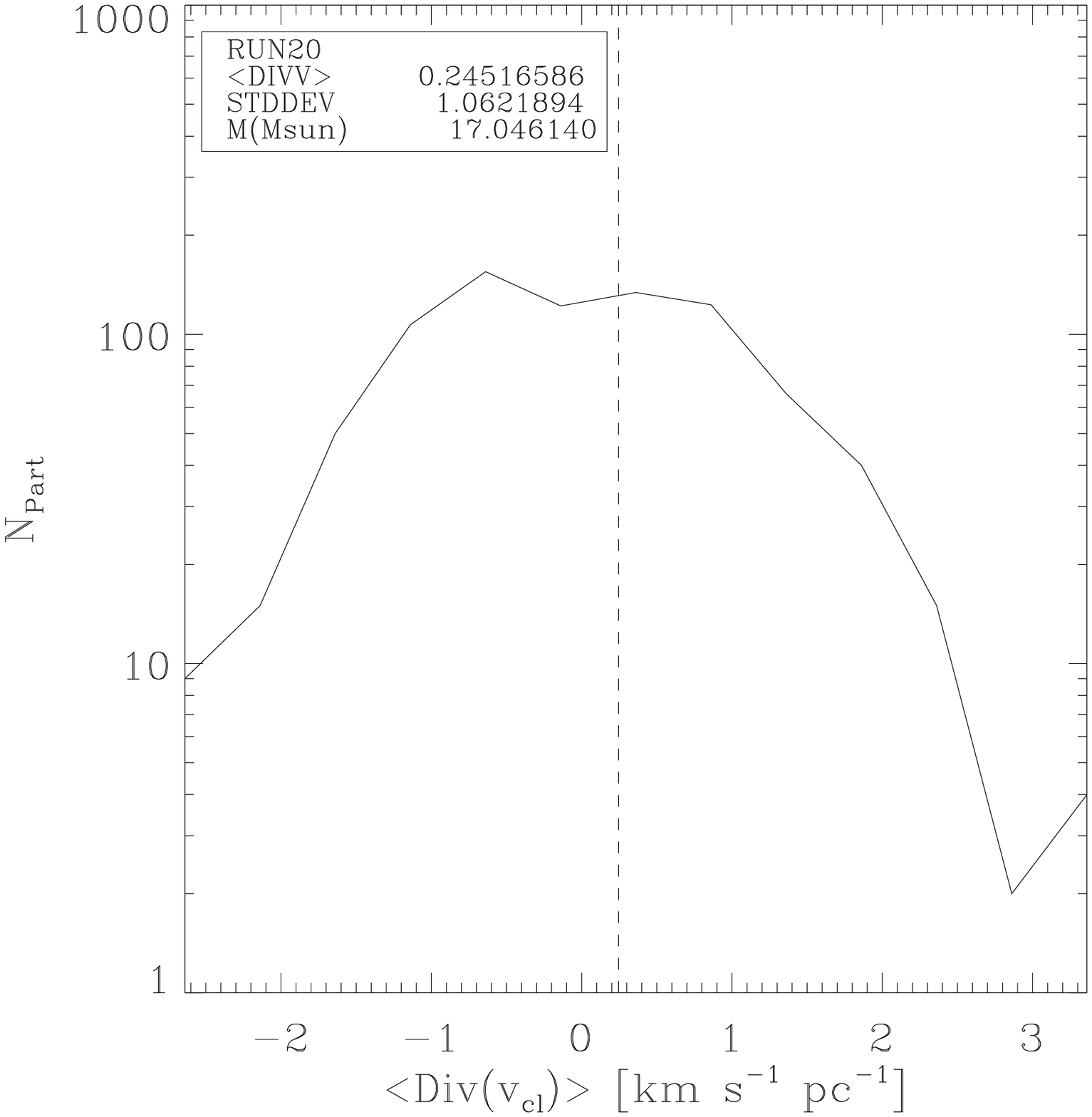}
\includegraphics[width=0.4\textwidth]{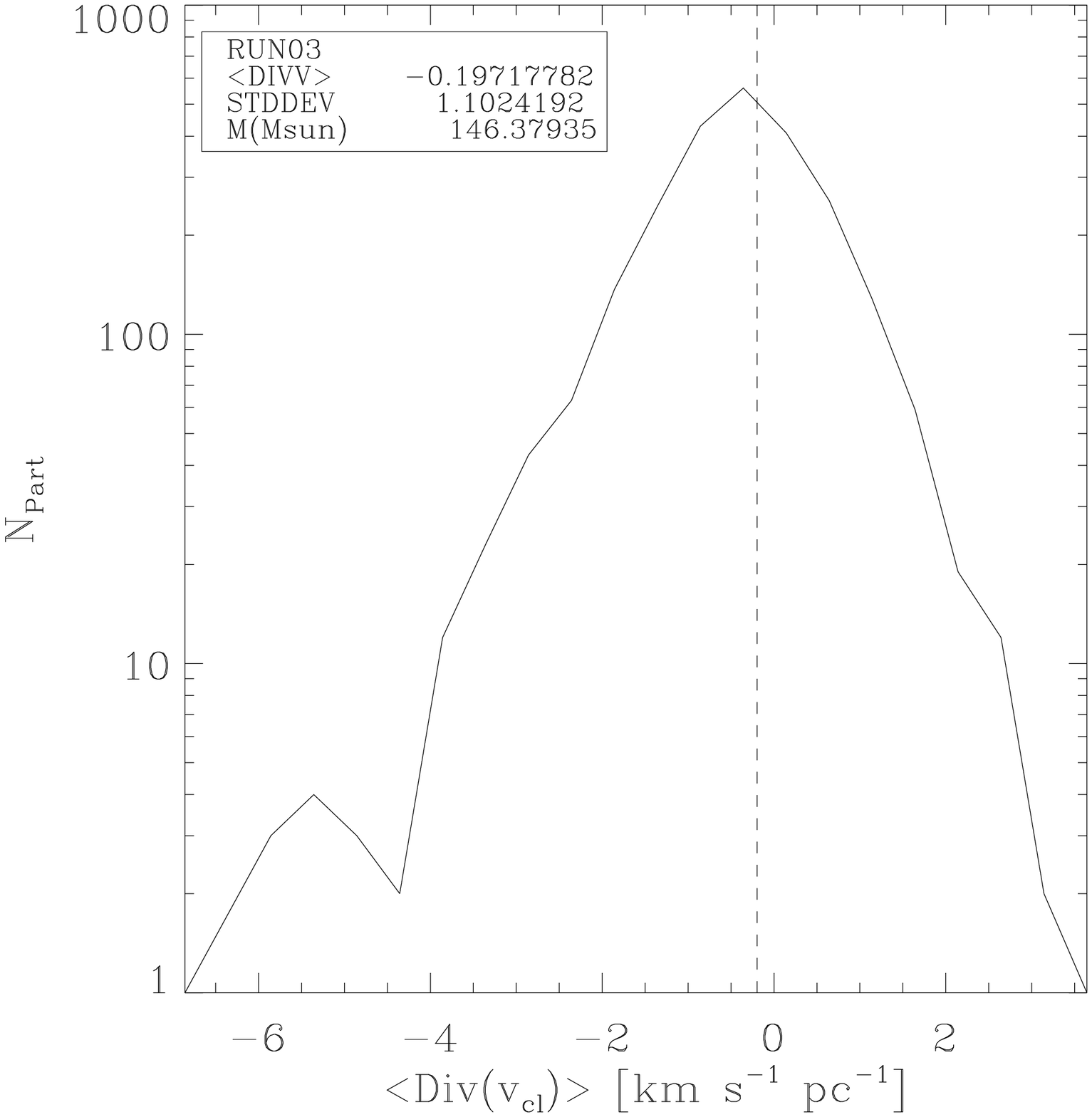}
\includegraphics[width=0.4\textwidth]{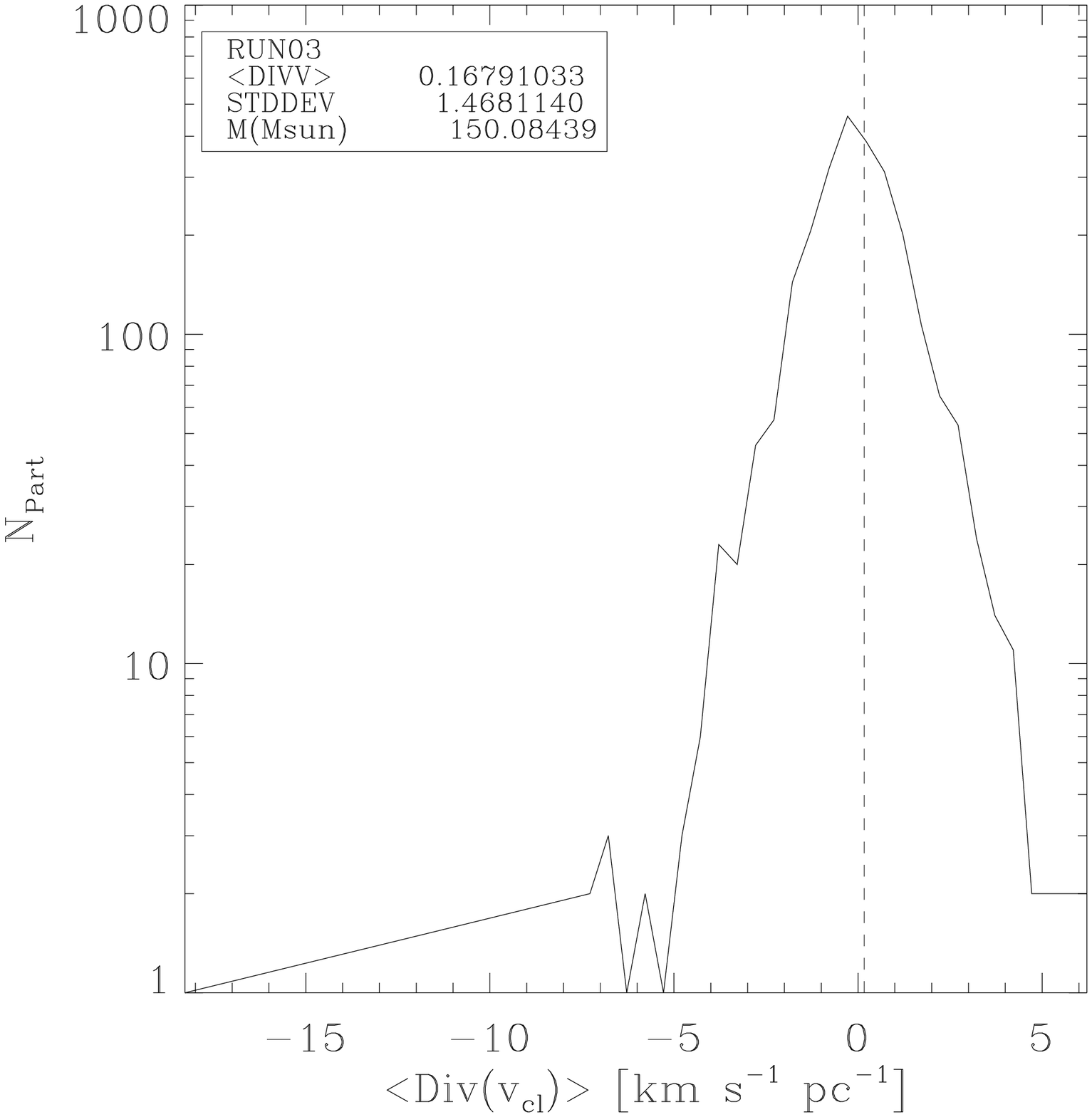}
\caption{Histograms of the velocity divergence for all the SPH particles
  belonging to four representative low- to intermediate-mass individual
  clumps from both simulations. The {\it top} row shows clumps from
  RUN20, and the {\it bottom} row shows clumps from RUN03. The {\it
    left} colums shows clumps with negative mean divergence (i.e,
  converging on average), while the {\it right} column shows clumps with
positive mean divergence (i.e., diverging on average).}
\label{fig:clump_divv}
\end{figure*}

Figure \ref{fig:clump_divv} shows the velocity divergence
histograms for all the SPH particles conforming some individual
clumps, at similar masses ({\it top}: low-mass, {\it bottom}:
intermediate-mass) and different values of the mean velocity divergence
({\it left}: negative, converging clumps; {\it right}: positive,
dispersing clumps). From this figure, we note that the clumps, in
general, contain a wide range of values of the
divergence; quite wider, in fact, than the range of {\it average}
divergences seen in Fig.\ \ref{fig:KH_low_nth}, as expected for the
distribution of partial averages of a random variable. In particular,
even the clumps with negative mean divergence contain a substantial
amount of particles where the local divergence is positive. Note,
however, that a local positive divergence does not necessarily imply
that the object is expanding globally. For example, a core undergoing
non-homologous collapse, with an increasing infall velocity towards
its center, will have a positive divergence in its envelope, because
of the stretching caused by the differential infall speed. This in
fact suggests that in fact our estimate of the fraction of contracting
clumps based on the mean divergence may actually underestimate the
actual fraction.

\subsubsection{High-column density clumps}
\label{sec:hi_col_dens}

While in the previous section we have discussed the kinetic energy
excess in low-column density clumps in the KH diagram, another 
feature in Fig.\ \ref{fig:Heyer} is that some of the highest-column
density objects also exhibit an excess of kinetic energy over the
equipartition value, especially in the case of \rv. 

As mentioned in Sec.\ \ref{sec:attempts}, clumps that appear
significantly above (by factors of a few) the virial equilibrium line in
the KH have often been interpreted as being gravitationally unbound,
and requiring an external confining pressure to avoid being dispersed in
a crossing time \citep [e.g.,] [] {KM86, Field2011}. However, it is also
possible that the apparent kinetic energy excess is due to an
underestimation of the relevant gravitational mass involved in the clump
dynamics, as proposed, for example, by H09 for their MC sample.  Two
mechanisms that come to mind for providing additional mass beyond that
directly measured in a clump are the mass in stars and the mass of
external accreting material that is part of the same gravitational potential 
well. Even if our high-$\Sigma$ clumps and cores do not show a very large 
excess in the $\sigmav/R^{1/2}$ ratio, we investigate their energetics 
under these two possibilities.

\subsubsubsection{The ``stellar mass effect''} 

\begin{figure*}[!t]
\centering
         \includegraphics[width=0.95\textwidth]{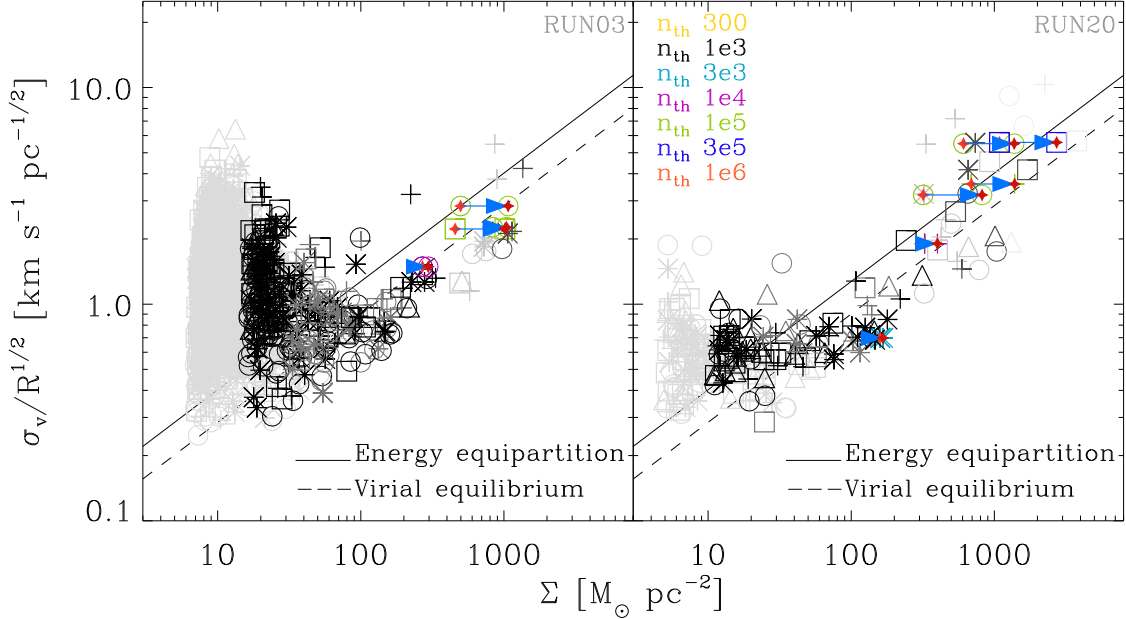}
         \caption{Correction to the location of sink-containing cores in
           the KH diagram (indicated by the arrows) due to the inclusion
           of the mass in sinks in the clumps' energy balance. 
           The cores with sinks are denoted by a red diamond. 
           As in figure \ref{fig:Heyer}, different symbols represent different times.}
      \label{fig:sink_eff}
\end{figure*}

One obvious source of mass in protostellar cores is the mass in 
(proto-)stellar objects which, in the case of cluster-forming clumps,
may reach observed values of up to 30--50\% \citep{LL03}. We thus
re-compute the location in the KH diagram of those cores and clumps
that do contain sink particles, adding the mass of the latter to the 
computation of the core column density ( i.\ e., now we consider 
$\Sigma = (M_{g}+M_{sink})/\pi R^2$ ). 
The result is shown in Fig.\ \ref{fig:sink_eff}, where symbols represent the
same timesteps  as in Fig. \ref{fig:Heyer} and clumps and cores that contain
stellar particles are marked with a red diamond.
From this figure, it is clear that the affected cores undergo a displacement in 
the KH diagram that relocates them closer to the virial equilibrium line in 
the case of RUN03, and to the region between the energy equipartition
and virial equilibrium lines in RUN20. However, some cores with such an 
excess do not contain sinks, and for them, the excess cannot be explained by 
this correction.

\subsubsubsection{The ``filament effect''} \label{subsec:filament}

A second possible mechanism for missing relevant gravitational mass in a
core may be if the core is gravitationally accreting material from a
surrounding structure, with the accretion driving turbulence into it 
\citep{KH10}. In particular, it has been found both in observations
\citep[e.g.,] [] {Schneider+10, Kirk+13, Peretto+13} and numerical
simulations \citep{Gomez2014}, that filaments may provide an accretion
channel of cloud material onto cores. In this case, it is reasonable to
ask whether the velocity dispersion in the core reflects the
gravitational potential of the entire filament/core system.

To test for this possibility, in both simulations we visually examined the
set of dense cores in our sample exhibiting a kinetic energy excess,
but not the ``sink effect'', to determine whether they belonged to a
filament. Rather surprisingly, we found none. We thus reversed the
procedure, visually searching for filament/core systems, and then
analyzing their energy budgets. 

\begin{figure*}[!t]
\centering
         \includegraphics[width=0.9\textwidth]{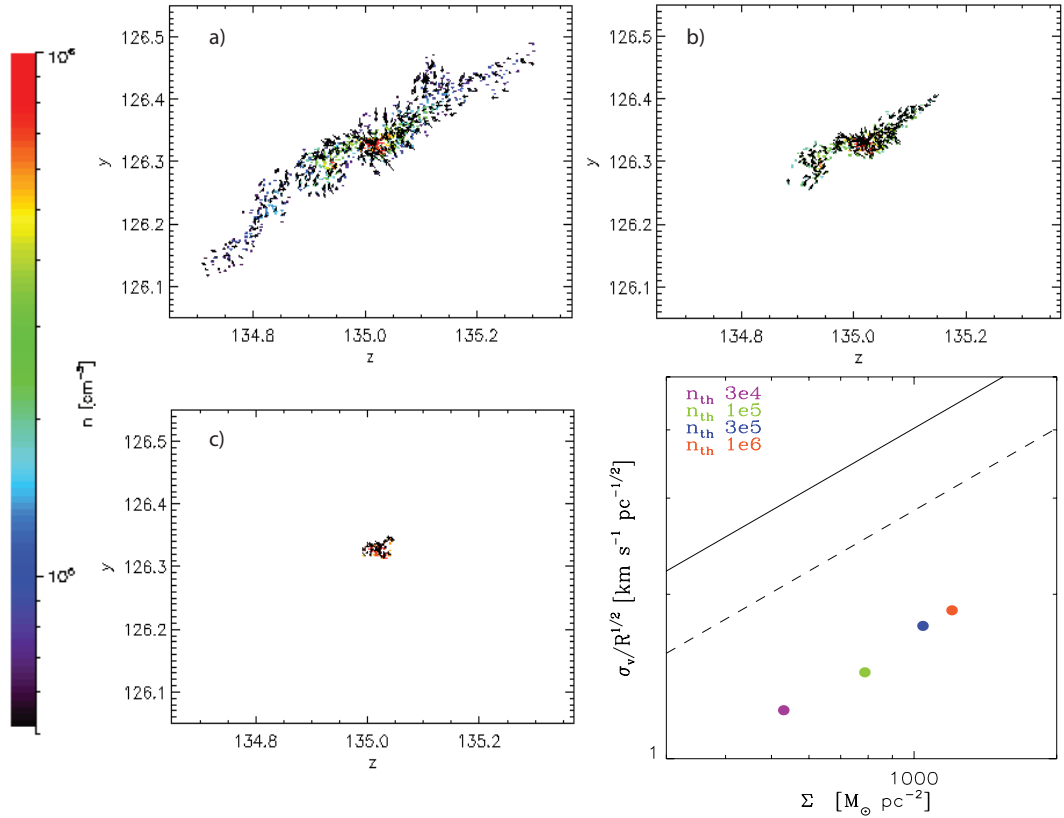}     
                  \caption{Filament/core system above  a) $\nth = 10^5$, 
                            b) $3\times10^5$ and c) $10^6\ \pcc$.
                            The right-bottom corner shows the 
                            filament/core system in the KH diagram at thresholds
                            $n_{th}=10^4$, $3 \times 10^4$, $10^5$, and $10^6\, \pcc$. 
            The points
           corresponding to the various thresholds are seen to describe
           a line parallel to the equipartition and virial-equilibrium
           lines, but displaced to a lower value of the ratio
           $\sigmav/R^{1/2}$, except for the point corresponding to
           $\nth = 10^6\, \pcc$, which is probably affected by numerical
          dissipation.}
      \label{fig:filament}
\end{figure*}

Figure \ref{fig:filament} shows one such filament/core system from 
RUN20 at $t = 26.5$ Myr at different density thresholds. The {\it right-
bottom} corner of this figure shows this system on the KH diagram.
Contrary to our expectation, this filament/core system exhibits a
{\it lower} value of the ratio $\sigmav/R^{1/2}$ than that expected for a 
spherical configuration, appearing {\it below} both the equipartition and 
the virial-equilibrium lines in this figure. In hindsight, this is actually 
natural, since the gravitational potential of a filamentary object of length 
$L$ is much lower than that of a spherical object of diameter $L$ and the 
same volume density, implying that the velocity dispersion of the former 
should be significantly lower than that of the latter
\citep{Toala+12, Pon+12}.

However, it might still be possible that the central, roundish core,
should have a larger velocity dispersion than that of an isolated core
of the same dimensions and in equipartition, because of the accretion
from the filament. Unfortunately, this system only appears roundish at
$\nth = 10^6\, \pcc$, and even at this threshold density, the core
appears sub-virial. In fact, we had to relax our minimum-mass selection
criterion (cf.\ Sec.\ \ref{sec:clump_sample}) in order to include this
core in the system, since it only contains 60 SPH particles, and it is
therefore likely to be significantly affected by numerical
dissipation. Thus, we cannot determine where it would be 
located in the KH diagram had it been free of numerical dissipation.

To try to answer this question, we searched for some other
filament/core systems at larger scale, so that the central core would
have a sufficient number of particles to be relatively free of numerical
dissipation. However, we have been unable to accomplish this task, because
of our restriction that the cores should have an SFE $< 65\%$ at $\nth =
10^5\, \pcc$. Indeed, we found that all larger cores that appeared to be
accreting from filaments already had efficiencies larger than this. This
seems to be a consequence of the fact that the filaments and the cores
grow roughly simultaneously, and accretion from the filament
onto the core only begins by the time the core has already undergone
significant sink formation, as also noted by
\citet{GO15}. This suggests that cores located within filaments in their
prestellar or early protostellar stages should not exhibit excess
kinetic energies, because they are not accreting significantly from
their filaments at these stages.

We conclude that filaments, and the pre- and protostellar
cores located within them, tend to exhibit sub-virial velocity
dispersions, due to the lower gravitational energies of these
configurations than those of the spherical structures assumed for the
virial velocity dispersion estimate. 

\subsubsubsection{Dispersing clumps and cores}

\begin{figure*}[!lt]
\centering
         \includegraphics[width=0.8\textwidth]{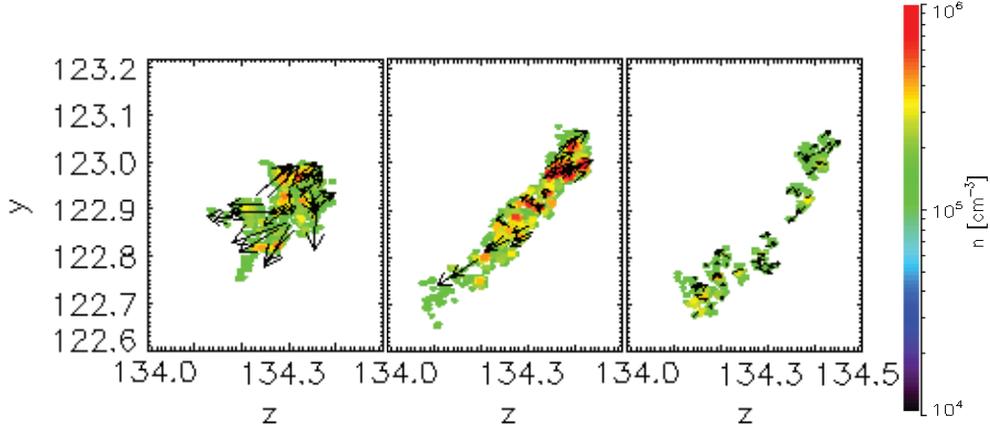}     
         \caption{High-column density core from RUN20 at times 
           $t =$ 22.2, 22.4 and 22.6 (from left to rigth )Myr at $n_{th}=10^5$cm$^{-3}$,
           showing that the core is being disrupted. The velocity field,
           shown by the arrows, also indicates that the clump is being
           dispersed, since it in general it is divergent.}
      \label{fig:disr_clump}
\end{figure*}

After considering both corrections by the mass in sink particles and
by existing in a filamentary environment, we are nevertheless left with
some dense cores whose kinetic energy excess cannot be explained by
either of these effects. Such is the case, for example, of the cores
indicated by the green and red `$+$' symbols with high values of
$\sigmav/R^{1/2}$ in the right panel of Fig.\ \ref{fig:sink_eff}. Figure
\ref{fig:disr_clump} shows this core at a threshold $\nth = 10^5\,
\pcc$ and at three times separated by 0.2 Myr, with the arrows
indicating the velocity field on the plane 
shown. It can be seen that the core is actually being disrupted, and so
this is indeed a case of a starless core that will probably never form
stars.

\subsubsection{Scatter propagation from KH to Larson} \label{sec:scatter}

Figure \ref{fig:Heyer} shows a
considerable scatter, especially for low-$\Sigma$. In  particular, it
is clear that the scatter is reduced as $\Sigma$
increases. We have interpreted this effect as a consequence of
 an increasing relative importance of
self-gravity at increasing column density, except for those high-density
objects that are being disrupted. Since we have argued that a
Larson-like linewidth-size relation appears for objects near
equipartition that are furthermore selected by near-constant column
density, the scatter around equipartition should cause a scatter around
the velocity dispersion-size relation as well. Indeed, a large 
scatter is also observed in the velocity
dispersion-size plots for both runs (cf.\ Fig. \ref{fig:Larson})

\begin{figure*}[!lth]
\centering
         \includegraphics[width=0.9\textwidth]{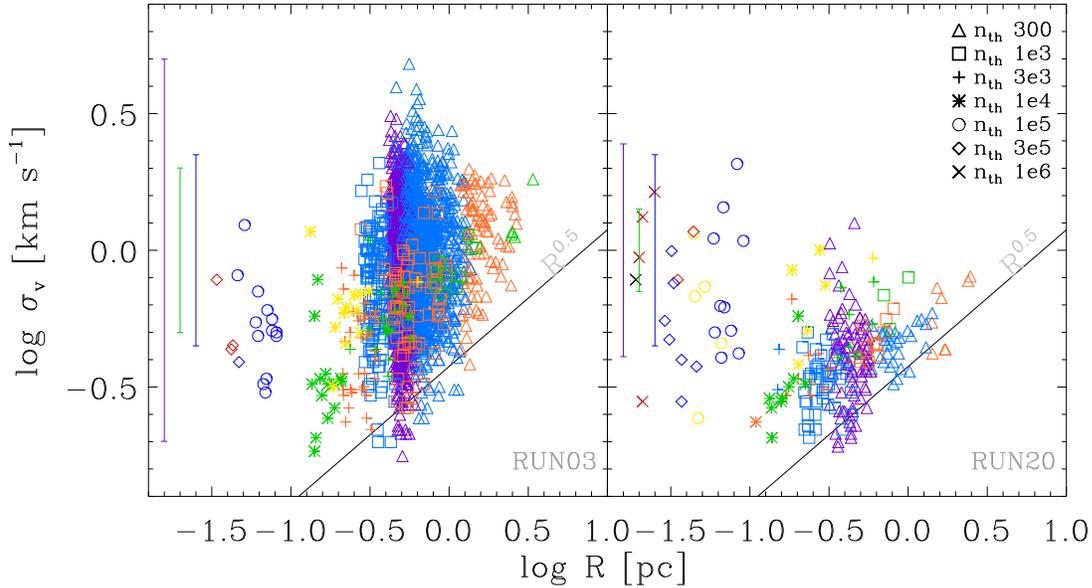}     
         \caption{Larson velocity-size relation for clumps and cores in both simulations.
         Colors and symbols are the same as in Fig. \ref{fig:Larson}. The error bars correspond
         to the scatter shown in Fig. \ref{fig:Heyer}, and their colors correspond to the $\Sigma$ range.
         }
      \label{fig:LarsonErr}
\end{figure*}

To quantify this, we note that the scatter in the quantity $\calh
\equiv \sigmav/R^{1/2}$ is related to that in the velocity dispersion
and in the radius by
\begin{equation}
d \ln \calh = d \ln \sigmav - \frac{1}{2} d \ln R.
\label{eq:scatter}
\end{equation}
The scatter $d\calh$ around the equipartition value $\left(2 \pi G
\Sigma/5\right)^{1/2}$ (cf.\ eq.\ [\ref{eq:free-fall}]) represents the
(physical) deviation from equipartition for a given clump.
Because the scatter $d \calh$ merges the scatter in $\sigmav$ and in
$R$, it is not possible to determine how $d \ln \calh$ is distributed
among $d \ln \sigmav$ and $d \ln R$. However, we can obtain an upper
limit in the expected scatter in $\sigmav$ if we assume that it
``absorbs'' all of the scatter in $\calh$, with none of it going to
$R$; that is, assuming $d \ln \sigmav = d \ln \calh$. In
Fig.\ \ref{fig:LarsonErr} we have  
plotted the error  bars for $\sigmav$ in the velocity
dispersion-size relation corresponding to the scatter in the
$\calh$ ratio from Fig. \ref{fig:Heyer},
for the three different
column density ranges (represented with the
color of the error bars, which are the
same as in the plots of Fig.\ \ref{fig:Larson}). 
It is clear from Fig.\
\ref{fig:LarsonErr} that 
the scatter in the Larson-like velocity
dispersion-size relation at low densities (purple points and error
bar) is clearly contained within the estimated upper limit originating
from the scatter in the KH diagram. Instead, for intermediate column
densities (green points and error bar), for which there is still a large
enough number of points to obtain good statistics and the scatter in the
KH diagram is not large, we see that the upper limit to the scatter
expected for $\sigmav$ is relatively small, and the points define a
clear Larson-like linewidth-size relation. We conclude that the
suggestion that clumps describe Larson-like relations when they are
restricted to narrow column density ranges and they are close to energy
equipartition is supported by our numerical clump sample.

\section{Discussion} \label{sec:discussion}

\subsection{Limitations} \label{sec:caveats}

As mentioned in Sec.\ \ref{sec:this_work}, our numerical simulations 
have a number of limitations. In particular, our SPH simulations neglect all 
forms of stellar feedback and magnetic fields. We plan to perform a similar 
analysis in a future contribution including these physical agents, but our
present study allows a first approximation to the problem. Furthermore,
our chosen setups only produce objects similar to observed low-mass
star-forming clumps and cores.

The neglect of stellar feedback has allowed us to investigate the energy
budget of clumps due exclusively to the interaction of initial background 
turbulence and self-gravity, without complicating the velocity field with 
additional contributions from the feedback. Our study has shown that in this 
context, our simulated clump sample reproduces observed trends in the KH 
diagram. In our sample, gravity has an increasingly dominant role as the 
column density of the clumps increases. On the other hand, low-column 
density clumps are increasingly dominated by turbulence, although large-scale 
turbulent compressions are dominant in more than half of these objects, in 
which external turbulent compressions provide the initial ``push'' that triggers 
the assembly of the clumps. In these, gravitational contraction is expected to 
take over when the clump has grown sufficiently massive. The other half may
actually correspond to clumps that will not grow to high densities and
masses.

Magnetic fields, on the other hand, even if insufficiently strong to support the 
clouds, as they are presently believed to generally be \citep[e.g.,] [] {Crutcher12}, 
might possibly delay the collapses \citep{Ostriker+99}, or reduce the number or 
formation rate of collapsing objects \citep{VS+05b, NL07}. This may affect the 
kinetic energies observed in the clumps, and we plan to repeat the present
analysis in a future contribution including both feedback and magnetic fields. 
Nevertheless, the similarity of the distribution of clumps in our simulations in the 
KH diagram to the observed one suggests that these agents may only play a 
secondary role during the assembly and early stages of collapse of clumps and 
cores.

\subsection{Applicability to low- and high-mass regions}
\label{sec:extrapol}

Strictly speaking, our results only apply to low-mass clumps and cores, since 
our sample does not include high-mass, high-column density (high-$\Sigma$) 
objects (see, e.g., the bottom panels of Fig.\ \ref{fig:Fall}) similar to cluster-
forming clumps, such as those studied by \citet{Fall+10}. Nevertheless, 
because massive-star forming cores do appear to follow the same scaling in the 
KH  \citep[e.g.,] [Ballesteros-Paredes et al. 2016, in preparation] {Ballesteros2011}
diagram as MCs and the low-mass cores we have discussed in Sec. \ref{subsec:heyer_alpha}, 
we speculate that our results may apply to high-mass, high-$\Sigma$ clumps. 
The clumps examined by \citet{Fall+10}, which were selected for their star
formation activity (not for their volume or surface density) have masses in
the range $10^2\lesssim M \lesssim 10^6 \Msun$ and surface densities 
$\Sigma \sim 10^3 \Msun$pc$^{-2}$. They are currently experiencing strong
stellar feedback  that may couneract with their own self-gravity.
The fact that  these clumps have roughly constant column density may indicate 
that these values of $\Sigma$ are physically selected by the requirement of 
exhibiting strong star formation and feedback. 
Column densities much higher than $\sim 10^3 \Msun$pc$^{-2}$ 
may not be observed because at that point gas removal from the clumps 
becomes important \citep{Fall+10}. On the other hand, at column densities 
lower than those values, the star formation activity may not be so strong 
\citep[][]{ZA+12, ZV14}, and the association between the gas and the stars may 
not be one-to-one because of the longer collapse timescales involved at lower
densities \citep{BH13}, and thus lower-column density objects will not
be selected by a strong star formation activity criterion.

\subsection{Comparison with previous work} \label{sec:compar_prev}

Our results can be compared with those obtained in two recent papers
that have performed numerical simulations of supernova (SN)-driven
turbulence in the ISM, one in a (250 pc)$^3$ cubic box \citep[]
[hereafter P+16] {Padoan+16}, the other in a parallelepiped-shaped box
of $1 \times 1 \times 40$ kpc$^3$, with a vertically stratified medium
\citep[] [hereafter IM+16] {Ibanez+16}. In both cases, turbulence is
driven for some time before turning on self-gravity. These two papers
have arrived at opposite conclusions concerning the distribution of the
simulated clouds in the KH diagram: IM+16 find that the MC-like objects
in their simulation develop near-equipartition after turning on
self-gravity, while P+16 find that their clouds never approach
equipartition, and instead have a roughly constant value of the ratio
$\calh \equiv \sigmav/R^{1/2}$, independent of the column density $\Sigma$,
suggesting that Larson's linewidth-size relation is valid after all, and
equipartition is not.  They also show that a sample of outer-Galaxy
clouds \citep{Heyer+01} appears to be consistent with this result.

The results from our initially random-driven, and subsequently decaying
simulations support those of IM+16 but not those of P+16,
since our clouds and clumps in general approach the equipartition
state. Moreover, we find that equipartition is more tightly fulfilled at
higher column densities, while low-column density objects exhibit larger
kinetic energy excesses over equipartition, in agreement with
observational data from various observational surveys \citep[e.g.,] []
{Barnes+11, Leroy+15}. Nevertheless, the most massive of the low-column
density objects are the ones closest to equipartition, suggesting that
the motions are dominated by gravity.

The origin of the contrasting results between IM+16 and P+16, both from
their respective simulations and from the observations each group
considered, deserves further examination. Concerning the simulations,
ample discussion has been given by IM+16, and here we just point out
that the simulations of P+16 may suffer from significant over-driving of
the turbulence. This is because they apply a standard value of the SN
surface density rate ($\sim 100$ Myr$^{-1}$ kpc$^{-2}$) into their (250
pc)$^3$ numerical box.  Although indeed most SN explosions are expected
to occur within this vertical span around the Galactic midplane, the
short height of the simulation box, which uses periodic boundaries, does
not allow the energy injected by the SNe to escape to high altitudes,
and to drive a galactic fountain, as it is known to do. Instead, this
energy must remain within the small volume of the simulation, likely
overdriving the turbulence in comparison with the actual observed levels
in the ISM. For example, Figure 6 in P+16 shows that the mean whole-box
velocity dispersion increases steadily from $\sim 20$ to $\sim 100\,
\kms$ during the last 10 Myr of evolution shown. However,
\citet[] [see also Gatto et al.\ 2015] {Scannapieco+12} have recently
found that in simulations where the total
velocity dispersion exceeds $\sim 35\ \kms$, the medium goes into a
thermal runaway regime where the gas is shocked into an
unstable regime in which the cooling time increases strongly with
temperature, causing a substantial fraction of the ISM to be unable to
cool on a turbulence dissipation timescale. As a consequence, the medium
goes into runaway heating, causing ejection of gas from any stratified
medium. Since the simulation by P+16 lacks such stratification, the
simulation is probably just heating up, explaining the continuous rise
of the velocity dispersion, and justifying our interpretation that this
simulation is overdriven and therefore not very realistic for the
purpose of examining the energy budget of the clumps.

On the other hand, concerning the outer-Galaxy cloud data used by
P+16, it is important to remark that these clouds have in
general quite low column densities, in the range 10--100 $\Msun$
pc$^{-2}$. Thus, they are indeed in the column density range where our
simulations indicate that turbulence is still dominant (see Fig.\
\ref{fig:Heyer}), even if, as clouds grow, they may later transition to
being dominated by self-gravity. In fact, the outer-Galaxy sample has
been plotted by \citet[see Fig.\ 13 of] [] {Leroy+15} together with data
from several other surveys, and it can be seen that the outer-Galaxy
clouds have the lowest-column densities and the largest scatter in the
$\calh$ parameter of the whole dataset, as with our results for the
low-column density clouds in our simulations. Nevertheless, when one
considers the whole dataset, including in particular objects of
substantially larger column densities, the tendency towards
equipartition is recovered, as shown in Figure 13 of
\citet{Leroy+15} and our own Fig.\ \ref{fig:Heyer}. We therefore
conclude that both the simulations and the data considered by P+16 are
restricted to regimes where indeed turbulence is dominant (either by too
strong turbulence driving or a low column density of the clouds), but
that these do not represent the general trend in the Galactic ISM when a
wide range of column densities is considered.

\section{Summary and conclusions}
\label{sec:concls}

In this paper we have investigated the intrinsic (rather than derived
from synthetic observations) physical conditions of clumps and cores in
two SPH simulations of the formation and evolution of molecular clouds
formed by converging motions in the warm neutral medium (WNM).  The two
simulations attempt to span a range of  likely motions in this medium. 

In both simulations, once the dense clouds form, they soon
begin to contract gravitationally, and some time later (a few Myr) they
begin to form stars, as in the general scenario described by
\citet{VS+07}  and \citet{HH08}. Neither of the simulations includes
turbulence-driving stellar feedback nor magnetic fields, and so all of
the kinetic energy is either driven by gravity or is a residual of the
turbulent/compressive motions that initiated the formation of the
clouds. Within this context of globally contracting molecular clouds
(MCs), we have investigated whether the clumps within them follow the
Larson scaling relations, or their generalization, as proposed by H09
and B11. We have also investigated the physical
conditions in clumps that appear to have an excess of kinetic energy, in
an attempt to understand the physical processes that cause this apparent
over-virialization.

We created an ensemble of clumps in each simulation by defining 
clumps as connected sets of SPH particles above a certain density-threshold 
$\nth$, so that a single clump at a lower threshold may contain several clumps 
at a higher value of $\nth$. The objects defined at the highest thresholds
($\nth \ge 10^5\, \pcc$) are refereed to as ``cores''.

Our results and conclusions may be summarized as follows:

\begin{itemize}

\item The full ensemble of clouds, clumps and cores does not follow either of 
the Larson scaling relations, but mostly follows their generalization, as 
proposed by H09 and B11. Nevertheless, low column density clumps in 
particular exhibit a large scatter, with a significant fraction of the clumps 
having values of the $\calh \equiv \sigmav/R^{1/2}$ parameter of up to an 
order of magnitude larger than the virial value, similar to the situation in 
various observational studies. 

\item We noted that, as emphasized  by B11, the kinetic energy implied 
by free-falling motions is only a factor of $\sqrt{2}$ larger than that for
virial equilibrium. We therefore generically refer to this condition as ``energy 
equipartition''. 

\item In our simulations, the equipartition condition is due to
gravitational contraction, by construction.

\item The clumps defined at a single threshold $\nth$ do not exhibit 
density-size or velocity dispersion-size relations. Instead, the exhibit nearly 
constant volume density, in agreement with previous studies \citep{Ballesteros2002}.
However, ensembles of clumps that exhibit near-equipartition and that are 
selected by column density ranges, do exhibit Larson-like relations, suggesting 
that these relations are special cases of the more general equipartition 
condition. 

\item We find examples of clouds, clumps and cores that exhibit excess
   kinetic energies over the equipartition level at both low- and
   high-column densities. Low-column density clumps that exhibit this
   excess are the least massive, while the more massive ones are closer
   to equipartition. Moreover, for more than 50$\%$ of the
   low-density clumps with an $\calh$ excess in both simulations, the velocity
   field in the clouds appears to be convergent 
   (i.e. have negative net divergence). 
   This suggests an evolutionary process
    in which a turbulent compression initially dominates the kinetic
    energy and exceeds the gravitational energy of the forming cloud.
    However, as the cloud becomes denser and more massive, the
    gravitationally-driven velocity becomes dominant.  Also, this
    suggests that the observation of an excess kinetic energy does not
    necessarily imply that a clump will disperse or needs an external
    thermal confining pressure to avoid dispersal. The excess kinetic
    energy may simply reflect the initial compressive motions 
    within the clump. In this case, instead of
    {\it confinement} of the cores by thermal pressure, we have {\it
      assembly} by ram pressure.

  \item Some of the high-column density cores that exhibit
    kinetic energy excesses contain stellar particles that
    increase the total gravitational potential in the volume of the
    clump. When this stellar mass is added to the gas mass in the
    energy budget of the core, the
    gas+stars system returns to near equipartition.

\item Some high-column density clumps with kinetic energy excesses,
however, do not contain stellar particles, so that the the above
correction cannot be applied. Investigation of the velocity field in
these cases does show a rotating and/or expanding motion, so that
these objects  are in the process of
  being disrupted, and will not form stars. Because this process
  is occurring at high densities, the driver of these disrupting motions
is likely to be the turbulence generated by the large-scale collapse.

\item  We also investigated the possibility that excess kinetic energies
in high-column density cores might be due to the cores being located in
filamentary clumps, with net accretion from the filament onto the core,
so that the velocity dispersion in the cores might represent the
gravitational potential of the mass in the filament. However, this
mechanism does not seem to be operational. We find that the filaments
and their embedded cores begin their evolution roughly simultaneously,
accreting material from the cloud mostly perpendicularly to the
filament. Accretion from the filament onto the core begins later, when
the core has become more massive and has already started to form stars. 
Thus, cores that are actively accreting from their parent filaments are already 
in advanced star-forming stages, and do not correspond to pre- or early 
protostellar objects.

\end{itemize}

\acknowledgments
The numerical simulations were produced in the cluster acquired through
CONACYT grant 102488 to E.V.-S. 
JBP acknowledges financial support from UNAM-PAPIIT grant number IN110816.


\bibliographystyle{aastex}


\clearpage
\pagestyle{plain}

\end{document}